\documentclass[aps,amsmath,amssymb,floatfix,showpacs]{revtex4}
\usepackage{amsmath,amssymb,amsfonts,mathrsfs}
\usepackage{color}
\usepackage{epsfig}
\usepackage{psfrag}
\usepackage{graphicx}
\usepackage{color}

\newcommand\bea{\begin{eqnarray}}
\newcommand\eea{\end{eqnarray}}
\newcommand\beq{\begin{equation}}
\newcommand\eeq{\end{equation}}
\newcommand\bib{\bibitem}

\def\non{\nonumber}
\def\dag{\dagger}
\def\al{\alpha}
\def\om{\omega}
\def\de{\delta}
\def\De{\Delta}

\def\ga{\gamma}
\def\ka{\kappa}
\def\la{\lambda}
\def\si{\sigma}

\begin{document}

\title{Majorana Fermions in superconducting wires: effects of long-range
hopping, broken time-reversal symmetry and potential landscapes}

\author{Wade DeGottardi$^1$, Manisha Thakurathi$^2$, Smitha Vishveshwara$^3$,
and Diptiman Sen$^2$}
\affiliation{\small{
$^1$Materials Science Division, Argonne National Laboratory, Argonne,
Illinois 60439, USA \\
$^2$Centre for High Energy Physics, Indian Institute of Science, Bangalore
560 012, India \\
$^3$Department of Physics, University of Illinois at Urbana-Champaign,
1110 W. Green Street, Urbana, Illinois 61801, USA}}

\date{\today}

\begin{abstract}
We present a comprehensive study of two of the most experimentally relevant
extensions of Kitaev's spinless model of a 1D $p$-wave superconductor: those
involving (i) longer range hopping and superconductivity and (ii)
inhomogeneous potentials. We commence with a pedagogical review of
the spinless model and, as a means of characterizing topological phases exhibited
by the systems studied here, we introduce bulk topological invariants as
well as those derived from an explicit consideration of boundary modes.
 In time-reversal invariant systems, we find that the longer range hopping leads
to topological phases characterized by multiple Majorana modes. In particular, we
 investigate a spin model, which respects a duality and maps to a
fermionic model with multiple Majorana modes; we highlight the connection between
 these topological phases and the broken symmetry phases in the original spin model.
 In the presence of time-reversal symmetry breaking terms, we show that the topological
 phase diagram is characterized by an extended gapless regime. For the case of
 inhomogeneous potentials, we explore phase diagrams of periodic, quasiperiodic,
 and disordered systems. We present a detailed mapping between normal state
 localization properties of such systems and the topological phases of the corresponding
 superconducting systems. This powerful tool allows us to leverage the analyses of
 Hofstadter's butterfly and the vast literature on Anderson localization to the question
 of Majorana modes in superconducting quasiperiodic and disordered systems, respectively.
 We briefly touch upon the synergistic effects that can be expected in cases where
 long-range hopping and disorder are both present.
\end{abstract}

\maketitle

\section{Introduction}

The recent explosion of studies concerning Majorana fermions in solid state
systems has brought the one-dimensional spinless $p$-wave paired
superconducting wire into the limelight. As the prototype for hosting
topological phases characterized by bound Majorana states at the ends of the
wire, this superconducting system has been theoretically studied from a
variety of angles~\cite{kitaev1,beenakker,lutchyn1,oreg,fidkowski2,potter,
fulga,stanescu1,tewari,gibertini,lim,tezuka,egger,ganga,sela,lobos,lutchyn2,
cook,pedrocchi,sticlet,jose,klinovaja,alicea,stanescu2,halfmetal,shivamoggi,
adagideli,sau,akhmerov,degottardi,ladder,niu,period,lang,brouwer2,brouwer3,vish}
and has formed the basis
for several experimental realizations~\cite{kouwenhoven,deng,rokhinson,das}.
While these analyses have led to revisiting several theoretical aspects
investigated in the past decade, such as topological features and symmetry
classification of the system, localization properties in the presence of
disorder, and physics in the presence of multiple channels, the prospect of
experimental realization has instigated new exploration of physical
realizations. A number of studies have focused on novel materials and
geometries, as well as subjecting the system to controlled external potentials.
Additionally, new means of detecting Majorana modes and signatures of related
non-Abelian statistics have been proposed, as well as on applications, such as
schemes for topological quantum computation in these
systems~\cite{tjunction,topqubit}.

Here, contributing to this vast literature, we present a comprehensive study
of the superconducting wire subject to various experimentally relevant
modifications of Kitaev's original model. By exploring several variants of the
coupling/hopping amplitudes and spatially varying electronic potentials, we
build on and unify previously studied features of similar models.
We investigate physics stemming from long-range hopping wherein conduction
electrons in a lattice version of the wire can hop across several sites.
In systems obeying time-reversal symmetry, we illustrate how such hopping
may give rise to multiple end Majorana fermions. We perform an involved study
of the wire in the presence of various potential landscapes, in particular,
periodic potentials, quasiperiodic potentials and disorder. Finally, we
briefly describe some of the richness associated with systems exhibiting both
long-range hopping and inhomogeneous potentials. Our goal here is to present
several new results of experimental relevance as well as to gather
foundational information that is scattered in the literature, recasting
some of it in simpler language and providing the non-expert a pedagogical,
self-contained exposition leading up to the results.

We begin with a review of the lattice version of the one-dimensional spinless
$p$-wave superconducting fermionic system, namely the Kitaev chain, and focus on its Majorana mode
properties. Specifically, we consider the phase diagram and topological features of the
system by virtue of the presence of Majorana end modes in some regions of parameter
space (topological phases) versus their absence in others (non-topological phase). We
 develop the formalism for several alternate topological invariants that are useful
in different circumstances. The relevant topological invariants can either be derived from bulk properties
for a homogeneous system with periodic boundary conditions or boundary properties
in long finite-sized wires which pinpoint the existence of end modes; the latter
was developed in our previous work in Ref.~\cite{ladder} within a transfer matrix formalism
and is easy to apply to a range of situations. This connection provides an elementary illustration of the so-called bulk-boundary correspondence~\cite{hasan,qi}. The topological invariants (TIs) are either $\mathbb{Z}_2$
in nature, detecting odd versus even number of Majorana end modes (and associated
 $\mathbb{Z}_2$ bulk physics), or $\mathbb{Z}$, detecting the presence of multiple,
 independent Majorana end modes (and associated winding numbers in the bulk), depending on whether time-reversal symmetry (TRS) is obeyed~\cite{hasan,qi,teo}.
It has been long known that the superconducting system can be mapped to a
spin chain, the topological phase being associated with an Ising ferromagnet and
the non-topological phase with a paramagnet~\cite{kitaev1,niu,ladder}; we recapitulate this mapping.

Detailed studies involving symmetry classifications of superconductors have identified the one
at hand as having a topological invariant that lies in $\mathbb{Z}$ when time-reversal
symmetry is preserved (class BDI). Here, we show that long-range hopping explicitly brings
out this $\mathbb{Z}$ character~\cite{fidkowski1}. Depending on the strength of the hopping terms, the
system can exhibit a slew of phases characterized by the presence of multiple, robust,
 independent Majorana end modes. In terms of bulk properties, these phases are
characterized by winding numbers having different integer values. While in terms
of lattice connectivity, the basic structure of these Majorana modes is the same as previously
discussed modes in multichannel wires~\cite{fidkowski1,potter,law}, our construction explicitly discusses the physics
of long-range hopping and can be realized within a single-channel system.

As a specific instance of a long-range hopping, we present a model having second nearest-neighbor
hopping that also displays interesting physics in terms of spin variables. The model exhibits four distinct
phases, three of which also have analogs in the Kitaev chain, and the fourth phase, being the novel
one, supports two independent Majorana modes at each end of a superconducting wire. In the
spin language, the model has cubic terms and an elegant duality map between one set of
spin variables and another. The four phases correspond to long-range ordering of different spin
components in the two sets of spin variables. The models shows a rich phase diagram highlighted
by the multiple-Majorana phase in the fermion language and by ordering involving duality maps
in the spin language.

The symmetry classification scheme has shown that in the absence of time-reversal symmetry,
the topological invariant associated with the system lies in $\mathbb{Z}_2$ (class D)~\cite{hasan,qi,teo,altland,brouwer1,gruzberg,fidkowski1}. Here,
by explicitly describing the system in terms of Majorana fermion degrees of freedom,
we discuss the manner in which the $\mathbb{Z}$ form of the invariant for time-reversal
symmetric systems gets reduced to $\mathbb{Z}_2$ once symmetry breaking terms are
introduced. We consider the effect of such terms on the Kitaev chain phase diagram, in
particular, the presence of a complex phase in the superconducting order parameter. We
show that such a phase gives rise to an unusual extended, gapless, capsule-like region
that lies between the gapped topological and non-topological regions.

The presence of spatially varying potentials too causes dramatic changes to the topological
phase diagram of the superconducting wire. Expanding on our results
of Ref.~\cite{degottardi}, we develop our transfer matrix formalism and show that
the manner in which Majorana end mode wave functions decay into the bulk of the
system can be connected to localization properties of a normal system (vanishing
superconducting gap) described by the same potential landscape (similar methods have been employed in~\cite{adagideli,beenakker,sau}). Moreover,
 for a common class of potentials, we show that there exists a mapping between the phase boundary
for fixed gap strength and its inverse. Armed with the transfer matrix tool, we analyze
the effect of several different potential landscapes. As the simplest case, periodic
potentials significantly affect the topological phase boundary, providing a knob
to control the extent of the topological regime. For quasiperiodic potentials, the
map to normal systems provides significant insight and reveals that the topological
phase diagram mirrors fractal-like structures, like the Hofstadter butterfly,
that naturally emerge in normal systems possessing quasiperiodicity.

Finally, for disordered potentials, the mapping to normal state properties is powerful in
that it allows us to leverage the extensive literature on Anderson localization in normal systems
 to identify the topological phase diagram for the disordered superconductor. We consider
a variety of potentials, including uniformly- and Lorentzian-distributed disorder.
All examples show that the topological phase continues to occupy a significant
region of the phase diagram in the presence of disorder. One of our findings is a
ubiquitous singularity in the phase boundary at the random-field transverse Ising
critical point. While we present a fairly extensive set of phase diagrams and
analyses of Majorana physics, our main contribution, the map to normal systems,
 is much more far-reaching in extent.

Our presentation is as follows. In Sec. I, we introduce our framework in the context of
a review of the Kitaev chain. In Sec. II, we describe the generic long-range hopping model
and Sec. III the specific instance of the four-phase model. In Sec. IV, we discuss the
effect of time-reversal symmetry breaking. In Sec. V, in the context of spatially varying
potentials, we further develop the transfer matrix technique introduced in Sec. I and
apply the methods to the case of periodic and quasiperiodic potentials. We analyze disordered
potentials in Sec. VI and conclude in Sec. VII.

\section{Review: Topological Aspects of a $p$-wave Superconducting Wire}

\label{sec:review}

We review the salient features of a single-channel $p$-wave paired superconducting wire within the context of
the broadly-used 1D tight-binding system of spinless electrons pioneered by
Kitaev~\cite{kitaev1} (referred to as Kitaev chain). We approach this simple and
well-studied system from various angles as a preparation for the new material in subsequent sections.
We discuss the phase diagram of the Kitaev chain in terms of its topological properties
characterized by the presence or absence of zero energy Majorana modes
at the ends of a long and open chain. To establish these topological
properties, we consider two symmetry classes, one respecting time-reversal invariance
and the other breaking it. To analyze these classes, we present various
 topological invariants, some exploiting bulk features of the Hamiltonian in
momentum space and others explicitly counting the number of
zero energy modes at the ends of an open chain. Finally, we review the
Jordan-Wigner transformation~\cite{lieb} which maps
the fermionic system to a spin-1/2 chain.

\subsection{Kitaev Chain: Model, Dispersion and Phases}

\label{sec:kitchain}

In this lattice description of the single-channel $p$-wave superconductor, electrons experience
a nearest-neighbor hopping amplitude $w$, a superconducting gap function for pairing between
 neighboring sites, $\De$, and an on-site chemical potential $\mu$; in this section, we will assume that
all these parameters are real. For a finite and open wire with $\mathcal{N}$ sites, the Hamiltonian
takes the form
\bea H = \sum_{n = 1}^{\mathcal{N}-1} \Big[ - w \left(f_n^\dag
f_{n+1} + f_{n+1}^\dag f_n \right) + \De \Big( f_n f_{n+1}
+ f_{n+1}^\dag f_n^\dag \ \Big) \Big] ~-~ \sum_{n = 1}^{\mathcal{N}}
\mu \left(f_n^\dag f_n - 1/2 \right), \label{ham0} \eea
where the operators $f_n$ satisfy the usual anticommutation
relations $\{ f_m, f_n \} =0$ and $\{ f_m, f_n^\dag \} = \de_{mn}$.
We can assume that $w>0$; if $w<0$, we can change its sign by the unitary
transformation $f_n \to (-1)^n f_n$. (Throughout this paper we will set both
$\hbar$ and the lattice spacing equal to unity). It is important to note that Majorana end
modes can only appear if the superconducting order parameter $\De \ne 0$.
This is because Majorana modes do not have a definite
fermion number, while the Hamiltonian commutes with the total fermion number,
$\sum_{n=1}^{\mathcal{N}} f_n^\dagger f_n$, if $\De = 0$

Towards exploring the Majorana mode structure of the wire, we can decompose the electron
operator in terms of real (Majorana) operators $a$ and $b$ as
\beq f_n ~=~ \frac{1}{2} ~(a_n + i b_n), ~~~~~~ f_n^\dag ~=~ \frac{1}{2} ~
(a_n - i b_n). \label{ab} \eeq
The operators $a_n$ and $b_n$ are Hermitian and satisfy $\{ a_m, a_n \} =
\{ b_m, b_n \} = 2 \de_{mn}$ and $\{ a_m, b_n \} = 0$.
Then Eq.~(\ref{ham0}) can be re-written as
\bea H &=& - \frac{i}{2} ~\sum_{n = 1}^{\mathcal{N}-1} \Big[ (w - \De) a_n b_{n+1}
+ (w+\De) a_{n+1} b_n \Bigl] ~-\frac{i}{2}~ \sum_{n = 1}^{\mathcal{N}}
\mu a_n b_n, \label{hamab}\eea

To study the bulk features of the system, we consider a long wire having
periodic boundary conditions (so that the first summation in Eq.~\eqref{ham0}
goes from $n=1$ to $\mathcal{N}$). Then
the momentum $k$ is a good quantum number and it goes from $-\pi$ to $\pi$
in steps of $2\pi/\mathcal{N}$. Defining the Fourier transform
 $f_k ~=~ \frac{1}{\sqrt \mathcal{N}} ~\sum_{n=1}^\mathcal{N} ~f_n ~e^{ikn}$, Eq.~\eqref{ham0}
can be re-written in momentum space as
\bea H &=& \sum_{0<k<\pi} ~\left( \begin{array}{cc}
f_k^\dag & f_{-k} \end{array} \right) ~h_k ~\left( \begin{array}{c}
f_k \\
f_{-k}^\dag \end{array} \right), \non \\
h_k &=& -(2w \cos k + \mu) ~s^z ~+~ 2 \De \sin k ~s^y, \label{hk1} \eea
where the $s^a$ are Pauli matrices denoting pseudo-spin degrees of freedom
formed by the fermion particle-hole subspace. The dispersion
relation follows from this and is given by
\beq E_k ~=~ \pm ~\sqrt{(2w \cos k + \mu)^2 ~+~ 4 \De^2 \sin^2 k}.
\label{energy} \eeq
The energy vanishes at certain values of $k$; these are given by lines
in the two-dimensional space of the parameters $\mu/w$ and $\De/w$. These
gapless lines correspond to phase transition lines which separate
different phases. The phase diagram consists of three lines demarcating
phases I, II and III, as shown in the top left diagram in
Fig.~\ref{fig:tr_broken_phase}. On the vertical red lines lying along
$\mu /w = \pm 2$, the energy vanishes at $k=\pi$ and zero respectively,
while on the horizontal blue line extending from $\mu/w = -2$ to $2$
at $\De/w = 0$, the energy vanishes at $k= \cos^{-1} (-\mu/(2w))$.

\begin{figure}[htb]
\begin{center}
\epsfig{figure=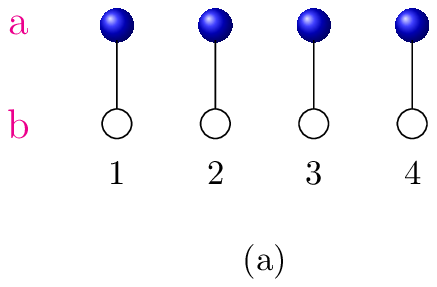,width=3.3cm} \hspace*{1.2cm}
\epsfig{figure=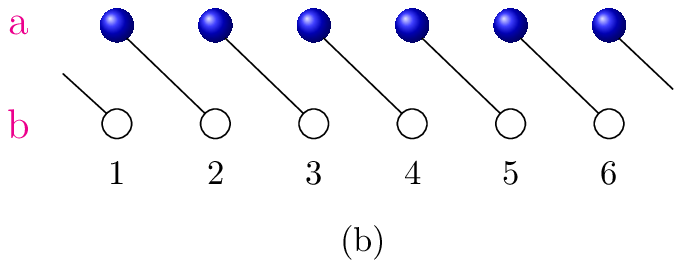,width=5.5cm} \hspace*{1.2cm}
\epsfig{figure=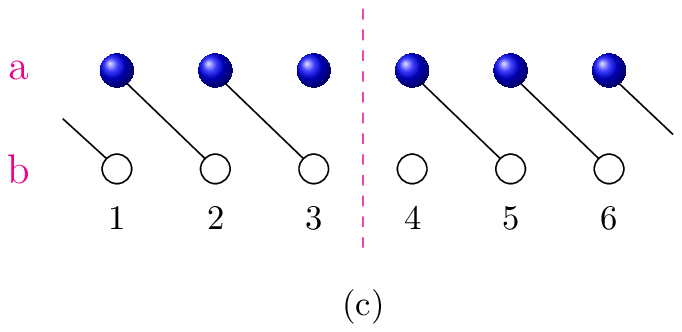,width=5.5cm} \\
\epsfig{figure=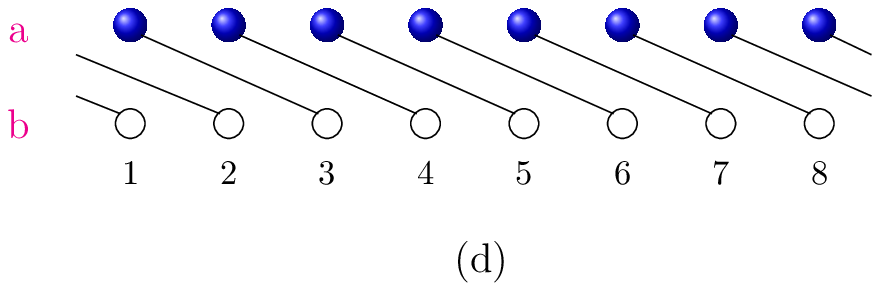,width=7cm} \hspace*{1.5cm}
\epsfig{figure=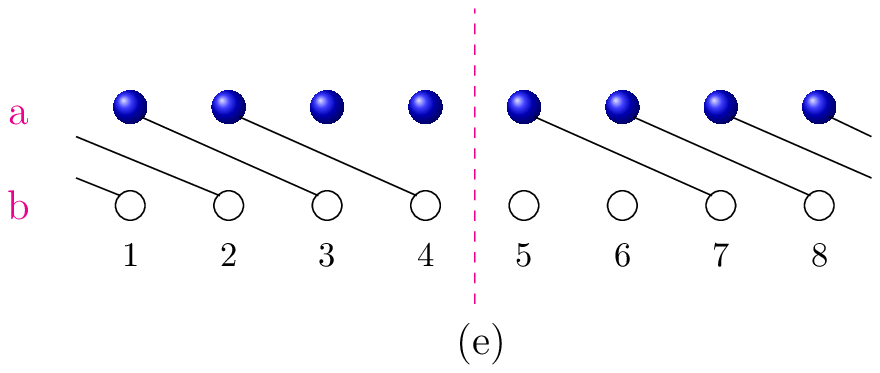,width=7cm}
\end{center}
\caption{Pictures of a fermionic chain showing the couplings
between Majorana operators $a_n$ and $b_n$
at different sites when only one kind of interaction, $a_m b_n$, is present
in the Hamiltonians in Eqs.~\eqref{hamab} and \eqref{ham4}. Panels (a), (b)
and (d) respectively show couplings of the form $a_n b_n$, $ a_n b_{n+1}$
(Eq.~\eqref{hamab}) and a longer range interaction $a_n b_{n+2}$
(Eq.~\eqref{ham4}) in the middle of a long chain. Panels (c) and (e) show
couplings of the form $a_n b_{n+1}$ and $a_n b_{n+2}$ for a chain which has
been cut into two, thus giving two open chains. Panel (c) has a Majorana mode
$a_3$ at the right end and a mode $b_4$ at the left end of an open chain.
Panel (e) has two Majorana modes $a_3$ and $a_4$ at the right end of an open
chain and two modes $b_5$ and $b_6$ at the left end of an open chain.
(Panel (a) would not have any Majorana end modes for a chain cut into two).}
\label{fig:endmodes} \end{figure}

\textbf{Phase Diagram}\--- To obtain insight into the nature of the phases, we
can consider some extreme limits. As shown in Fig.~\ref{fig:endmodes} (a),
for $\De=w=0$, $\mu\neq 0$, which lies in phase III of the phase diagram,
we see in terms of the Majorana mode Hamiltonian of Eq.~\eqref{hamab} that
each Majorana mode $a_n$ on a given site is bound to its partner $b_n$ with
strength $\mu$, leaving no unbound modes. For $\De=w\neq 0$, $\mu= 0$,
lying in phase I, the only existing bonds connect $a_n$ to its neighbor
$b_{n+1}$ (Fig.~\ref{fig:endmodes} (b)), leaving
a free $a/b$-Majorana mode at the right/left end of a finite sized system
(Fig.~\ref{fig:endmodes} (c)). For $\De=-w\neq 0$, $\mu= 0$, lying in
phase II, the roles of $a$ and $b$ modes become interchanged. As shown using
topological arguments in the next section, the presence/absence of these end
modes is robust in that deviations from these extreme limits in parameter
space does not change these features unless a phase boundary associated
with a vanishing gap is crossed.

\subsection{Topological Invariants (TIs), Transfer Matrix Approach and
Majorana Modes}

The topological properties of the superconducting wire described above can be captured in a variety of ways, all involving global features of the system characterized by TIs. There are several TIs developed in the literature; some are a type of generalized winding number (for instance, the celebrated TKKN invariant of the integer quantum Hall effect~\cite{thouless}) or are based on a Pfaffian, as in~\cite{kitaev1}. Building on previous work~\cite{ladder,degottardi}, we introduce complementary `boundary invariants', i.e. TIs which are derived from an explicit counting of the number of edge Majorana states. It should be emphasized that although the bulk and boundary invariants are derived in different ways, they all share two crucial characteristics: (1) they are restricted to integer values and (2) small deformations of the Hamiltonian which do not close a bulk gap cannot change their values. Furthermore, they enumerate the number of edge Majorana modes. The different forms will
be used in subsequent sections based on convenience and the aspects studied.

\textbf{Symmetry Classes}\--- In this paper we will encounter both $\mathbb{Z}$ and $\mathbb{Z}_2$ TIs (which we explain below). These types indicate the level of topological protection enjoyed by a topological insulator
or superconductor. An extensive classification of topological insulators and superconductors has been developed, and the type of TI can be determined from the class which the Hamiltonian falls into~\cite{hasan,qi,teo}. The class depends on a number of factors, such as whether the particles are bosons or fermions and whether the
Hamiltonian has spin rotation symmetry (if the particles have a non-zero
spin), particle-hole symmetry or TRS. The different symmetry classes have been enumerated and there is extensive discussion in the literature~\cite{altland,motrunich,brouwer1,gruzberg,fidkowski1}. Relevant to this work are the classes:

{\bf Class D} -- This class is appropriate for describing spinless electrons exhibiting $p$-wave pairing but for which time-reversal symmetry is broken~\cite{altland,brouwer1,gruzberg,fidkowski1}. This class exhibits $\mathbb{Z}_2$ topological protection (in 1D) indicating that the \emph{parity} of the number of Majorana modes is protected~\cite{kitaev1,hasan,qi,teo}.

{\bf Class BDI} -- This class also describes spinless electrons exhibiting $p$-wave pairing, with the additional restriction that time-reversal symmetry is obeyed. This class is associated with a $\mathbb{Z}$
TI in 1D, indicating that the number of Majorana modes itself is topologically protected~\cite{fidkowski1}.

As is seen here, a general feature of the `periodic table of topological insulators' is that the level of topological protection is generally greater as the symmetry constraints on the Hamiltonian become
more restrictive~\cite{hasan,qi,teo,fidkowski1}.

\subsubsection{Bulk Invariants}

Depending on the time-reversal properties of the Hamiltonian $h_k$,
two different invariants can be defined. (Although Eq.~\eqref{hk1}
defines $h_k$ only for $0 \le k \le \pi$, it will now be convenient to use
the same expression for $h_k$ for the entire range $0 \le k \le 2 \pi$).
In Eq.~\eqref{hk1}, we see that $h_k$ is of the form $h_k = a_{3k} s^z +
a_{2k} s^y$, which maps to the vector $\vec V_k = a_{3k} \hat z + a_{2k}
\hat y$ in the $z-y$ plane.
The form is general to this class of systems which have TRS,
i.e., when $h^*_{-k} = h_k$ for all $k$. Our first bulk
invariant $W$, as introduced in Refs.~\cite{niu,tong}, is then the winding
number associated with the angle $\phi_k = \tan^{-1} (a_{2k}/a_{3k})$ made
by the vector $\vec V_k $ with respect
to the $\hat z$ axis upon traversing the Brillouin zone, i.e.,
\beq W ~=~\int_0^{2\pi} ~\frac{dk}{2\pi} ~\frac{d\phi_k}{dk}.
\label{wind} \eeq
This object can take any integer value and is a TI, namely, it
does not change under small changes in $h_k$ unless $h_k$ happens to pass
through zero for some value of $k$; in the latter case, the winding number
becomes ill-defined and the energy $E_k = 0$ which means that the bulk gap
closes at that value of $k$. As will be explicitly discussed for the case of long-range hopping, we thus have a $\mathbb{Z}$-valued invariant, as expected from the general classification discussed above. It is straightforward to determine the relationship between $W$ and the number of edge modes $N_a$ and $N_b$ of $a$ and $b$ type (respectively) of Majorana modes at the left-hand end of an open chain, and confirm that $W = N_b - N_a$.
Note that taking $\De \rightarrow - \De$ reverses the winding of $\vec{V}_k$ and thus $W$, which is consistent with this transformation interchanging $a$ and $b$ Majoranas (see Sec.~\ref{sec:kitchain}).

In the absence of TRS, i.e., if $h^*_{-k} \ne h_k$
for some $k$, as seen in later sections, $h_k$ generally has four components,
$h_k = a_{0k} I + a_{1k} s^x +
a_{2k} s^y + a_{3k} s^z$, and it is not possible to define a winding
number for the corresponding vector in four dimensions. On the other hand,
for momentum values $k=0$ and $\pi$, $h_k$ has only one component, namely,
$h_0 = h(0) s^z$ and $h_\pi = h(\pi) s^z$. As seen in Eq.~\eqref{hk1}, this
is naturally true also in the case with TRS. Under the
stringent assumption that the system is fully gapped (i.e., gapped for
all values of $k$), one can define a $\mathbb{Z}_2$-valued TI:
\beq \nu_{bulk} = \mbox{sgn} \left( h(0) h(\pi) \right).
\label{eq:bulkZ2top} \eeq
This invariant can take the values $-1$ or $1$ and is topological in that its
value cannot change unless either $h(0)$ or $h(\pi)$ crosses zero in which
case the energy $E_0$ or $E_\pi$ vanishes. We will see that $\nu_{bulk}$
is equal to the parity of the number of
Majorana modes at the end of the system.

The fact that systems which break TRS have a symmetry only at the level $\mathbb{Z}_2$ may also be seen another way. As seen above, sending $\De \rightarrow - \De$ takes $W \rightarrow -W$. This process can be carried out for class D \emph{without} closing a gap since we can take $\De = \De_0 e^{i \varphi}$ and take $\varphi$ from $0 \rightarrow \pi$. However, for class BDI, $\De$ is constrained to be real and the only way to take $\De \rightarrow - \De$ is to pass through zero, which closes a gap~\cite{fidkowski1}.

\subsubsection{Boundary Invariants: Transfer Matrix Approach}
\label{sec:boundaryinvariants}

We now outline the transfer matrix approach detailed in our previous work,
Ref.~\cite{ladder}, for identifying the Majorana modes structure at the end of a wire.
It will become apparent that the transfer matrix
explicitly gives us $N_a$ and $N_b$ (as defined in the previous section, $N_{a,b}$ is the number of $a$- and $b$-type Majorana modes at the left-hand side of the system, respectively). We also note that finding $N_a$ and $N_b$ from the transfer matrix
will only involve finding the eigenvalues of a matrix; this is numerically
easier than calculating than $W$ which involves doing
an integral over the momentum $k$.

The transfer matrix can be obtained from the Heisenberg equations of motion
for the Majorana operators in Eq.~\eqref{hamab}:
\bea i \frac{da_n}{dt} &=& - ~[H, a_n] ~=~ -i (w+\De) b_{n+1} ~-~ i (w-\De)
b_{n-1} ~-~ i\mu b_n, \non \\
i \frac{db_n}{dt} &=& - ~[H, b_n] ~=~ i (w+\De) a_{n-1} ~+~ i (w-\De) a_{n+1}~+~
i\mu a_n. \label{eom1} \eea

Assuming that these operators depend on time as $a_n = \al_n e^{-iEt}$
and $b_n = \beta_n e^{-iEt}$, we find
the values of the energy $E$ for which the above equations have solutions. The
solutions in the bulk have the same dispersion as the one given in Eq.~\eqref{energy};
 in particular, the energies differ from zero by a finite gap except on the phase boundaries.
In addition to these bulk modes, in the topological phases there are end modes which lie at
 zero energy for long chains and are therefore separated by a gap from the bulk modes.
(For short chains, modes at the two ends can hybridize and lift their degeneracy away from zero.)
 For $E=0$, Eqs.~\eqref{eom1} take the form
\bea (w+\De) \al_{n+1} ~+~ (w-\De) \al_{n-1}~+~ \mu \al_n &=& 0, \non \\
(w+\De) \beta_{n-1} ~+~ (w-\De) \beta_{n+1} ~+~ \mu \beta_n &=& 0.
\label{eom2} \eea

These equations can be represented in the transfer matrix form
\beq \label{eq:matrixA}
\left( \begin{array}{c}
\al_{n+1} \\
\al_n \end{array} \right) =~A_n \left( \begin{array}{c}
\al_n \\
\al_{n-1} \end{array} \right),~~\mbox{where}~~
A_n = \left( \begin{array}{cc}
-\frac{\mu}{\De + w} & \frac{\De - w}{\De + w} \\
1 & 0 \end{array} \right). \eeq
Since the $A_n$ may be taken as functions of $\mu/w$ and $\De/w$,
we set $w = 1$. A similar expression holds for the transfer matrix $B_n$
for the $\beta_n$. (In a later section, we will consider models where the
chemical potential $\mu_n$ and therefore $A_n$ and $B_n$ vary with $n$).

\begin{figure}[htb]
\begin{center}
\epsfig{figure=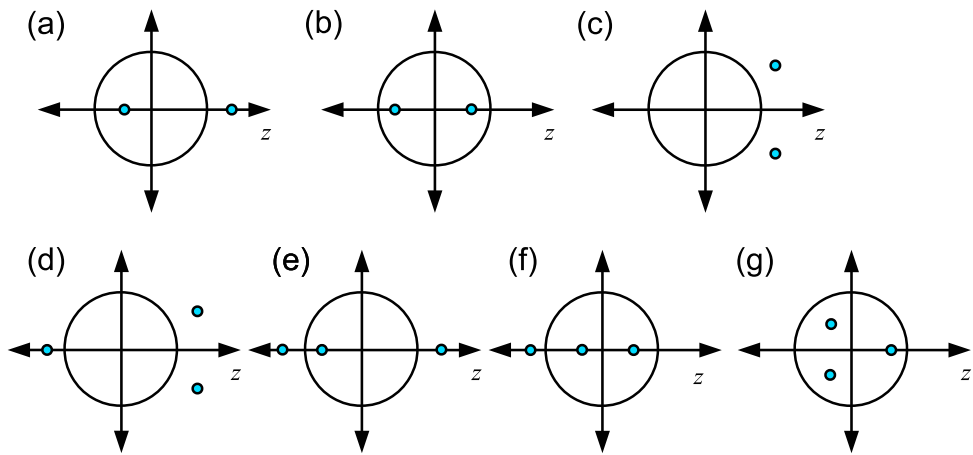,width=13cm}
\end{center}
\caption[Eigenvalues of the transfer matrix.]{Panels (a-c): Possible locations
of the eigenvalues of a $2 \times 2$ transfer matrix for Majorana mode $a$ as
discussed in Sec. II B. Only one eigenvalue may lie inside the unit circle as
in (a), or both eigenvalues may lie inside the unit circle as in (b), or both
may lie outside the unit circle as in (c). Panel (d-g): Possible locations of
the eigenvalues of a $3 \times 3$ transfer matrix for Majorana mode $b$ as
analyzed in Sec. V A. As discussed in the text, the number and type of
Majorana modes at each end of a long chain is governed by the number of
eigenvalues inside the unit circle; cases (a) and (e) give non-topological
phases characterized by the absence of end modes, while all the other cases
give topological phases with one or more end modes at each end.}
\label{fig:eigenvalues} \end{figure}

The existence of end Majorana modes requires the $\al_n$ (or $\beta_n$) to
be normalizable, i.e., $\sum_n |\al_n|^2$ (or $\sum_n |\beta_n|^2$) should
be finite. The number of eigenvalues of the full transfer matrix
$\mathcal{A} \equiv \prod_{n = 1}^{\mathcal{N}} A_n$ with magnitude less
than 1 is denoted by $n_f$. The number $n_f$ corresponds to the number of roots
of the characteristic polynomial for the transfer matrix, $f(z) = \det
\left(I - \mathcal{A} z\right)$, that lie within the unit circle. Hence,
\bea n_f = \frac{1}{2 \pi i} \oint_{|z| = 1} ~dz~ f'(z)/f(z),
\label{eq:nf} \eea
as was noted in the context of plane wave zero modes in Ref.~\cite{wen}.
One can show that the eigenvalues of the full
transfer matrix $\mathcal{B} \equiv \prod_{n = 1}^{\mathcal{N}} B_n$ are
inverses of the eigenvalues of $\mathcal{A}$, and therefore the number of
eigenvalues of $\mathcal{B}$ with magnitude smaller than 1 is $2-n_f$.
For $n_f = 0$ and 2, $\al_n$ and $\beta_n$ are normalizable and therefore
the system is topological (with a $\al$ mode at one end and a $\beta$ mode
at the other end of a long chain), whereas for $n_f = 1$, $\al_n$ and
$\beta_n$ are not normalizable and the system is non-topological.

With $n_f$ at hand, we can define a $\mathbb{Z}_2$ invariant
\beq \nu_{end} ~=~ -~ (-1)^{n_f} = - \mbox{sgn} \left( f(1) f(-1) \right),
\label{eq:TIinvariant} \eeq
for which $\nu_{end}=-1$ and $\nu_{end}=1$ reflect an odd versus even number of
normalizable end Majorana modes, respectively. For the Kitaev chain of
Eq.~\eqref{ham0}, since there can only be one or zero modes, these
values correspond to topological
and non-topological phases, respectively. Since the topology of the system
depends only on the magnitude of $\De$, we take $\De$ to be positive;
hence $|\det~\mathcal{A}| < 1$. Then the two eigenvalues of $\mathcal{A}$
obey $|\la_1 \la_2| < 1$. Therefore, for $|\la_1|<|\la_2|$,
we have $|\la_1| < 1$ and $n_f$ is completely determined by the larger
eigenvalue $\la_2$. Thus, we have that $\nu = \mbox{sgn} \left( \ln |
\la_2 | \right)$ for $\De > 0$. We claim that the odd-even Majorana
structure denoted by the $\mathbb{Z}_2$ invariant $\nu_{end}$ of Eq.~\eqref{eq:TIinvariant}
matches the form of the bulk invariant $\nu_{bulk}$ of Eq.~\eqref{eq:bulkZ2top}.

\textbf{Phase Diagram} \--- We can now revisit the Kitaev chain phase diagram in
light of these invariants. Considering some special cases, for $\mu = 0$ and
$\De > 0$, we find that both eigenvalues of $\mathcal{A}$
have magnitude smaller than 1, so that $n_f = 2$. Further, we find that there
is a zero energy Majorana mode of type $a$ at the left end and of type $b$ at
the right end of a long chain. This describes phase I following the discussion
after Eq.~\eqref{energy}. If $\mu = 0$ and $\De < 0$, both eigenvalues
of $\mathcal{A}$ have magnitude larger than 1, so that $n_f = 0$. We find
that there is a zero energy Majorana mode of type $b$ at the left end and of
type $a$ at the right end of a long chain. This describes phase II. Finally,
let us consider $\De = 0$ and $|\mu| > 2t$. This lies in phase III. We then
find that $n_f = 0$ and that there are no Majorana modes at either end of
a long chain. This is summarized in Table~I.

\begin{table}[htb]
\begin{center}
\begin{tabular}{|c|c|c|c|c|c|}
\hline
Phase & $\nu$ & $W$ & $n_f$ & $N_a$ & $N_b$ \\
\hline
I & $-1$ & $-1$ & $2$ & $1$ & $0$ \\
II & $-1$ & $1$ & $0$ & $0$ & $1$ \\
\hline
III & $1$ & $0$ & $1$ & $0$ & $0$ \\
\hline
\end{tabular}
\caption{Summary of the topological phases (I and II) and non-topological
phase (III) exhibited by the Hamiltonian of Eq.~\eqref{ham0} and the different
topological invariants. Here $\nu$ refers to both $\nu_{bulk}$ and
$\nu_{end}$, consistent with their equivalence.} \end{center}
\label{tab:topdata} \end{table}

Note that on the phase transition lines which separate the various phases,
the bulk gap closes, i.e., $E_k = 0$ at certain real values of the momentum
$k$. Eqs.~\eqref{eom2} and \eqref{eq:matrixA} then imply that $A_n$ must have
an eigenvalue of the form $e^{ik}$ which lies on the unit circle. Thus the
eigenvalues of the transfer matrix cross the unit circle as we go across a
phase transition line.

The TIs that we have discussed here are related to
other invariants which exist in the literature. We refer the reader
to Refs.~\cite{teo,wen,ladder}.

\subsection{Mapping to a Spin-1/2 Chain}~\label{sec:spinmap}

The Hamiltonian in Eq.~(\ref{hamab}) can be mapped to that of a spin-1/2
chain; here we briefly review the mapping and associated physics~\cite{lieb}.
We define the Jordan-Wigner (JW) transformation between a
spin-1/2 and a spinless fermion at each site $n$ so that the states with
$\si_n^z = \pm 1$ correspond to the fermion number $f_n^\dag f_n = 1$ and
0 respectively. The JW transformation takes the form
\bea a_n &=& \left( \prod_{i=1}^{n-1} \si_i^z \right) ~(-1)^n ~\si_n^y, ~~~~~~
b_n ~=~ \left( \prod_{i=1}^{n-1} \si_i^z \right) ~(-1)^n ~\si_n^x.
\label{jw} \eea
Eq.~\eqref{hamab} can then be re-written as
\beq H ~=~ -~ \sum_{n = 1}^{\mathcal{N}-1} \Big[ J_x \si_n^x \si_{n+1}^x +
J_y \si_n^y \si_{n+1}^y \Bigl] ~-~ \frac{1}{2} ~\sum_{n = 1}^{\mathcal{N}} ~
h \si_n^z, \label{hamxyz} \eeq
with
\begin{equation}
J_x = \left(w - \De \right)/2,\ J_y = \left(w + \De\right)/2,\ h = \mu.
\end{equation}

The Hamiltonian describes a spin chain having nearest-neighbor $xx$ and $yy$
couplings and a magnetic field pointing in the $z$ direction.

In the discussion below Eq.~\eqref{energy}, we had stated that there are three
phases, I, II and III, depending on the values of the parameters $\mu /w$ and
$\De/w$. While in the fermionic language the ordering is topological in nature and
has no local order, we can see that in the spin language, the corresponding
phases are described by ordering of local spin variables. Assuming that $w >
0$, let us first consider $\mu = 0$. Then the point $\De = w$ lies in phase I
and corresponds to $J_x = 0$ and $J_y = w > 0$. This describes a ferromagnetic
Ising model with $yy$ couplings which has long-range order in $\si^y$.
Similarly, $\De = -w$ lies in phase II and describes a ferromagnetic
Ising model with $xx$ couplings which has long-range order in $\si^x$.
Finally, $\De = 0$ and $|\mu| \gg 2w$ describes a model with $J_x = J_y$
and a magnetic field in the $z$ direction whose magnitude is much larger
than $J_x$. This describes a system with no long-range order in either
$\si^x$ or $\si^y$. These special points represent the entire phase diagram
in that phase III is that of a disordered paramagnet, and phases I and II
have Ising ferromagnetic order (along $y$ and $x$, respectively) and are
separated by a gapless line ($\De=0$, $J_x=J_y$) describing an $xy$
ordered spin chain in a transverse field.

While the spin chain described above has a venerable history in and of itself,
the mapping sets the stage for more complicated spin systems and mappings
in the context of topological order, such as with the Kitaev honeycomb model
\cite{kitaev2} and the Kitaev ladder~\cite{vish,ladder}. In Sec. IV, we will
discuss an interesting generalization of the fermionic model, which has multiple
Majorana modes and exhibits a rich phase diagram, that can be elegantly
described in terms of spin variables.

\section{Long-range Hopping and Multiple Majorana End Modes}

Here we argue that the presence of longer range hopping that extends beyond
the nearest neighbor has the dramatic consequence that in a spinless
superconducting wire that preserves TRS, multiple topologically protected
Majorana modes can form at the ends of the wire. These modes are
stable in that regions in parameter space corresponding to different number
of modes are protected by bulk gaps in the energy spectrum
and correspond to topologically distinct phases. In the previous analyses of
symmetry classes mentioned above, it has been argued in various ways that
the TI lies in $\mathbb{Z}$ for the class currently being
considered, class BDI. The discussion here in terms of long-range hopping in
a single-channel model and of the TIs defined above gives
a simple, direct and comprehensive picture for the $\mathbb{Z}$ form by way
of multiple Majorana end modes.

Below we introduce the general long-range hopping model and analyze its topological properties
based on the presence of multiple Majorana end modes. In the next section,
we explore the features of the aforementioned model having four topologically distinct
phases and a dual Ising representation in the language of spins.

\subsection{Long-range Hopping Model}

As a very general case of long-range hopping, we consider a modification
to the Kitaev chain of the previous section that takes into account an
infinite set of couplings for long-range hopping. Focusing purely on the
hopping and pairing, this generalized version of the Majorana Hamiltonian of
Eq.~\eqref{hamab} for an infinite wire takes the form
\beq H ~=~ -i \sum_{r=-\infty}^\infty~ \sum_{n=-\infty}^\infty ~J_r ~a_n
b_{n+r}, \label{ham4} \eeq
where the $J_r$ are real parameters. Note that we have chosen to discuss
long-range hoppings in a single chain. However, our model is equivalent,
for appropriate choices of the $J_r$, to a multi-chain system with both
interchain and intrachain couplings.

The Hamiltonian in Eq.~(\ref{ham4}) is invariant under TRS which
involves complex conjugating all numbers, taking the time
$t \to - t$, changing $a_n \to - a_n$ and keeping $b_n$ unchanged. The
transformation of $a_n$ is justified by Eq.~(\ref{jw}) where we see that
complex conjugating reverses the sign of $\si_n^y$ and therefore of $a_n$.
Note that the square of the time-reversal transformation is equal to
$+1$ since it leaves $a_n$ and $b_n$ unchanged.

In terms of Dirac fermions $f_n = (a_n + ib_n)/2$ and $f_n^\dag =
(a_n - ib_n)/2$, the generalized momentum space version of Eq.~(\ref{hk1})
 momentum space stemming from Eq.~(\ref{ham4}) takes the form
\bea H &=& \sum_{0<k<\pi} ~\left( \begin{array}{cc}
f_k^\dag & f_{-k} \end{array} \right) ~h_k ~\left( \begin{array}{c}
f_k \\
f_{-k}^\dag \end{array} \right), \non \\
h_k &=& -2 \sum_{r=-\infty}^\infty ~[ J_r \cos (kr) ~s^z ~+~ J_r \sin (kr) ~
s^y], \label{ham5} \eea
where the $s^a$ are once again Pauli matrices.

Note that the Hamiltonian is time-reversal invariant: $h^*_{-k} = h_k$.
The energy-momentum dispersion follows from Eq.~(\ref{ham5}),
\beq E_k ~=~ \pm 2 \sqrt{\left(\sum_r J_r \cos (kr) \right)^2 ~+~
\left(\sum_r J_r \sin (kr) \right)^2}. \eeq
The phase diagram can, in principle, be found from this expression by
demanding that $E_k$ should vanish for some value of $k$ lying in the
range $[0,\pi]$ or by employing the TIs defined in the
previous section. While this is a difficult problem, we will see below that
a great deal of insight can be gained by considering situations in which one of
the $J_r$ is much larger than all the others.

\subsection{Multiple Majorana Modes and Topological Phases}

To address what phases are exhibited by the model for an infinitely
long chain and how many zero energy Majorana modes there are
for the ends of a long chain in the different phases, consider a Hamiltonian
where one of the $J_r$, say $J_q$ for some positive integer $q$, is non-zero
and positive while all the others are zero~\cite{fidkowski1}.
(From Eq.~(\ref{ham4}) we see that the ground state is then given by a state
in which $i a_n b_{n+q} = 1$ for each $n$). We show
this situation in Fig.~\ref{fig:endmodes} (d) for the case of $q=2$. In
Fig.~\ref{fig:endmodes} (e), we have shown a dotted line which cuts the chain
into two; it is clear that the open chain on the right side has $q$ Majorana
modes of type $b$ at its left end, while the open chain on the left side has
$q$ Majorana modes of type $a$ at its right end. Since we expect the number
of Majorana modes
and their types to be TIs, this phase will survive for
small changes in the values of all the other $J_r$. We therefore conclude
that the model in Eq.~(\ref{ham4}) has an infinite number of phases which
can be labeled by an integer $q$ which describes the number and types of
Majorana modes at the ends of a long chain.

Next, we look at what the $\mathbb{Z}$-valued bulk invariant, $W$,
defined in Eq.~\eqref{wind}
gives for the Hamiltonian in Eq.~\eqref{ham5}. As before, we can think of
$h_k$ as defining a two-dimensional vector. If only one of the couplings
$J_q$ is non-zero and positive, the vector is given by
\beq \vec V_k ~=~ - 2 ~[J_q ~\cos (kq) ~\hat z ~+~ J_q ~\sin (kq) ~\hat y].
\label{vk} \eeq
As $k$ goes from $0$ to $2\pi$, this generates a closed curve
which encircles the origin of the $z-y$ plane $q$ times in the clockwise
direction. Defining $\phi_k = \tan^{-1} (V_y/V_z)$ and using Eq.~\eqref{wind},
we find that the winding number is equal to $q$ for the configuration given
in Eq.~(\ref{vk}), i.e., $W=q$. Thus the winding number is also
equal to the number of Majorana modes of type $b$ ($a$) at the left (right)
end of a long chain. We can now consider what happens if all the $J_r$ are
allowed to be non-zero; the vector is then given by
\beq \vec V_k ~=~ - 2 \sum_r ~[J_r ~\cos (kr) ~\hat z ~+~ J_r ~\sin (kr) ~
\hat y]. \eeq
The winding number is a TI
and therefore does not change under small changes in all the $J_r$. The system
thus continues to remain in the phase $q$ as long as the closed curve
does not pass through the origin for any value of $k$. But if the curve
passes through the origin for some value of $k$, the energy vanishes at
that value of $k$ and the system lies at a quantum critical point
separating two phases having different topological values $q$.

\textbf{Transfer matrix approach with constraint equations} \--- We can also understand the existence of Majorana end modes using the
transfer matrix approach. Consider a semi-infinite chain, with
sites going from $n=1$ to $\infty$, with a Hamiltonian of the form
\beq H ~=~ \sum_{n=1}^\infty ~\sum_{m = 1}^q \left( w_m f_{n+m}^\dag f_n +
\De_m f_{n+m}^\dag f_{n}^\dag + \mbox{H.c.} \right) ~-~ \sum_{n=1}^\infty ~
\mu_n (f_n^\dag f_n - 1/2), \label{eq:genham} \eeq
where $w_m$, $\De_m$ and $\mu_n$ are all real, and $q \ge 1$ is an integer.
Let us now study the zero energy equations of motion to see if there are
Majorana modes localized near $n=1$, i.e., the left end of the chain.
To be specific, let us focus on the $a_n$ modes. Each site $n \ge q+1$ gives
rise to an equation linking $a_n$ to all sites from $a_{n-q}$ to $a_{n+q}$;
these equations can be described by a $d \times d$
transfer matrix $A$, where $d = 2q$. However, for $n=1,2,\cdots,q$, all
the sites up to $a_{n-q}$ are not present in the system; hence the
corresponding equations are not of the transfer matrix form, but
instead provide $q$ constraints on the first $2q$ values of $a_n$. Let us now
suppose that the parameters $(w_m,\De_m,\mu_m)$ are such that the transfer
matrix $A$ has $n_f$ eigenvalues with magnitude smaller than 1, and the other
$d-n_f$ eigenvalues have magnitude larger than 1. The eigenvectors
corresponding to the first $n_f$ eigenvalues are normalizable. However, the
presence of $q$ constraints means that there are only $n_f - q$
independent and normalizable end modes. Thus, the number of $a$ Majoranas on the left-hand
side of the system is
\begin{equation}
N_a = n_f - q.
\end{equation}
If $n_f \le q$, there are no Majorana end modes of type $a$. A similar analysis
carried out for the $b_n$ modes indeed shows that there are an equal number of $b$ Majoranas localized
to the opposite end of the system. Finally, we also note that this argument implies that $a$ and $b$ type Majoranas can never occur on the same side of the system. Such a state is incompatible with TRS. As illustrations, models of the
form given in Eq.~\eqref{eq:genham} with next-nearest neighbor hopping and
multiple Majorana modes are presented in Fig.~\ref{fig:multiple} as well as
in Sec. IV.

While the above analysis gives the exact number of end Majorana modes
of types $a$ and $b$, it can also be useful to find an expression for the
$\mathbb{Z}_2$-valued invariant analogous to Eq.~\eqref{eq:TIinvariant}.
We can derive this as follows. Since the transfer matrix
$\mathcal{A}$ is real, its eigenvalues must either be real or must come in
complex conjugate pairs. If we define $f(\la) = det (\mathcal{A} - \la I)$, we have
the relation $(-1)^{n_f} = \mbox{sgn} \left( f(1) f(-1) \right)$.
(This relation holds because if an eigenvalue of $\mathcal{A}$, denoted by
$\la_i$, is a real number not equal to $\pm 1$,
then $\la_i^2 - 1 < 0 ~(>0)$ depending on whether $\la_i$ is smaller than
(larger than) 1 in magnitude. If $\la_i$ is complex, then $(\la_i^2 - 1)(
\la_i^{*2} -1) > 0$). If we define
\beq \nu_q = (-1)^q \mbox{sgn} \left( f(1) f(-1) \right),
\label{eq:topinvgen} \eeq
we see that $\nu_q = +1 ~(-1)$ corresponds to having an even (odd)
number of Majorana modes of type $a$ at the left end of a chain. In
particular, $\nu_q = -1$ means that there is at least one Majorana end
mode and therefore the system is in a topological phase. Thus
Eq.~\eqref{eq:topinvgen} is the appropriate generalization of
$\nu_{end}$ (Eq.~\eqref{eq:TIinvariant}) to arbitrary values of $q$ and $n_f$

Finally, a few comments are in order on the general long-range hopping
model. We have seen that when only couplings between $a$ and $b$ modes are
present, the system has $\mathbb{Z}$ topological symmetry, allowing for an arbitrary
number of end Majorana modes. When TRS is broken,
however, intra-couplings between the $a$'s or the $b$'s themselves become
manifest and, as seen is following sections, this couples modes
near each end, allowing for only a $\mathbb{Z}_2$ symmetry. On another note, as for
mappings between the fermions and spin outlined in the previous section,
Eq.~\eqref{eq:genham} yields spin Hamiltonians typically containing
multi-spin terms involving arbitrarily long strings of $\si^z$. While
these general cases are too complex to provide further insight, we present
a model having a tractable mapping to a spin model in the next section.

\section{Long-range Hopping: Ising Duality and a Four-phase Model}

As an illustration of the long-range Majorana hopping model discussed in the
previous section, we now make a detailed study of a model with four
parameters and four associated phases. We present this model not only
as an instance of supporting multiple end Majorana modes but also as a novel
spin system that supports an Ising duality and interesting interpretations
of phases in terms of spin ordering. The spin Hamiltonian that we introduce
not only has usual linear and quadratic terms in the spin language, but also an unusual cubic term, and is of the form
\beq H ~=~ - ~\sum_n ~[ J_x \si_n^x \si_{n+1}^x + \mu \si_n^z + J_y
\si_n^y \si_{n+1}^y - \nu \si_{n-1}^x \si_n^z \si_{n+1}^x]. \label{ham1} \eeq
%This is invariant under the $Z_2$ symmetry $\si_n^x \to - \si_n^x$,
%$\si_n^y \to - \si_n^y$ and $\si_n^z \to \si_n^z$.
Taking the site label $n$ to run over all integers in Eq. (\ref{ham1}), we
can define a dual spin-1/2 chain whose site labels run over $n + 1/2$,
with the mappings
\bea \tau_{n+1/2}^z &=& \si_n^x \si_{n+1}^x, ~~~~~~
\tau_{n-1/2}^x \tau_{n+1/2}^x ~=~ \si_n^z, \non \\
\tau_{n-1/2}^x \tau_{n+1/2}^z \tau_{n+3/2}^x &=& - ~\si_n^y
\si_{n+1}^y, ~~~~~~ \tau_{n-1/2}^y \tau_{n+1/2}^y ~=~ - ~\si_{n-1}^x
\si_n^z \si_{n+1}^x.
\label{map} \eea
Eq. (\ref{ham1}) then becomes
\beq H ~=~ - ~\sum_n ~[ \mu \tau_{n-1/2}^x \tau_{n+1/2}^x + J_x
\tau_{n+1/2}^z + \nu \tau_{n-1/2}^y \tau_{n+1/2}^y - J_y \tau_{n-1/2}^x
\tau_{n+1/2}^z \tau_{n +3/2}^x], \label{ham2} \eeq
which interchanges $J_x \leftrightarrow \mu$ and $J_y \leftrightarrow \nu$
with respect to Eq. (\ref{ham1}); this is the duality property.
%Eq. (\ref{ham2}) is invariant under the $Z_2$ symmetry $\tau_{n+1/2}^x \to -
%\tau_{n+1/2}^x$, $\tau_{n+1/2}^y \to - \tau_{n+1/2}^y$ and $\tau_{n+1/2}^z
%\to \tau_{n+1/2}^z$.

We note here that a model with the Hamiltonian given in Eq.~(\ref{ham2})
but with only three parameters, $\mu$, $J_x$ and $J_y$, was elegantly
analyzed in Ref.~\cite{niu}. Since that study did not include the parameter
$\nu$, it did not enjoy complete duality; further, the transfer matrices
considered there were two-dimensional (as in our earlier sections)
rather than three-dimensional as we will discuss below.

As a map to a fermion model, we invoke the Jordan-Wigner
transformations of Eq.~\eqref{jw},
$a_n ~=~ \left( \prod_{i=-\infty}^{n-1} \si_i^z \right) ~(-1)^n ~\si_n^y$ and
$b_n ~=~ \left( \prod_{i=-\infty}^{n-1} \si_i^z \right) ~(-1)^n ~\si_n^x$, to
obtain the Hamiltonian
\beq H ~=~ -i ~\sum_n ~[\nu a_n b_{n+2} + J_x a_n b_{n+1} + \mu a_n b_n
+ J_y a_n b_{n-1}] \label{ham3} \eeq
In principle, we can perform a Jordan-Wigner transformation from the
$\tau$ variables as well, and the resultant Majorana Hamiltonian also
makes the duality manifest.

\subsection{Majorana Modes, Spin Ordering and Phases}

As in the previous section, we identify zero energy modes employing
Heisenberg equations of motion for the Majorana operators:
\bea \nu b_{n+2} + J_x b_{n+1} + \mu b_n + J_y b_{n-1} &=& 0, \non \\
J_y a_{n+1} + \mu a_n + J_x a_{n-1} + \nu a_{n-2} &=& 0, \eea
with similar equations for $d_{n+1/2}$ and $c_{n+1/2}$ respectively.
Taking $b_n = \la^n$, we obtain the cubic equation
\beq \nu \la^3 + J_x \la^2 + \mu \la + J_y = 0. \label{root} \eeq
This equation is essentially the characteristic polynomial equation $det
(B - \la I) = 0$ for the $3 \times 3$ transfer matrix $B$ which relates
$(b_{n+2}, b_{n+1}, b_n)$ to $(b_{n+1}, b_n, b_{n-1})$.

This polynomial equation has three roots, at least one of which must be real,
and the complex roots must come in complex conjugate pairs. (The corresponding
equation for $a_n = \ka^n$ is given by $J_y \ka^3 + \mu \ka^2 + J_x \ka +
\nu = 0$. The roots of this are clearly the inverses of the roots of
Eq.~(\ref{root}),
and it suffices to know the roots of one equation to deduce that of the other).
Note that if all the four parameters in Eq. (\ref{root}), $\nu, ~J_x, ~\mu, ~
J_y$, are scaled by the same factor, the roots remain the same. Further, just
as the three roots are completely determined by the values of the four
parameters, the four parameters are also completely determined, up to an
overall scale factor, by the value of the three roots: if the three roots
are $\la_1, ~\la_2, \la_3$, we must have
\beq (\la - \la_1) ~(\la - \la_2) ~ (\la - \la_3) ~=~ \la^3 + \frac{J_x}{\nu}
\la^2 + \frac{\mu}{\nu} \la + \frac{J_y}{\nu}. \label{inv} \eeq

Assuming that none of the roots $\la_i$ lie on the unit circle, there are
four possible cases as shown in Fig.~\ref{fig:eigenvalues} (d-g). By
considering certain extreme limits of the couplings, based on which spin
ordering dominates and on the transfer matrix structure for the $a$ and $b$
modes, we can characterize the four regions as follows.

\textbf{Phase A}\--- All three roots lie inside the unit circle.
This is the case if
$|\nu| ~\gg ~|J_x|, ~|\mu|, ~|J_y|$. Eq. (\ref{ham2}) then implies that the
$\mathbb{Z}_2$ symmetry $\tau_{n+1/2}^y \to - \tau_{n+1/2}^y$ is spontaneously broken
and $\tau_{n+1/2}^y$ develops long-range order. (If $\nu \to \infty$, we have
$+1$ or $-1$) at all sites, while if $\nu \to - \infty$, we have an
a ferromagnetic state in which $\tau_{n+1/2}^y$ has the same value (either
antiferromagnetic state in which $\tau_{n+1/2}^y$ takes the values $+1$ and
$-1$ on alternate sites. In either case, $\tau_{n+1/2}^y$ has long-range
order). As for the end modes, we can use the arguments based on the transfer
matrix approach with constraint equations in Sec. IV B to show that this phase
has two Majorana modes of type $b$ at the left end and two of type $a$ at the
right end.

\textbf{Phase B}\--- Two of the roots lie inside the unit circle
while one lies outside. This is the case if $|J_x| ~\gg ~|\nu|, ~|\mu|, ~|
J_y|$. Eq. (\ref{ham1}) then implies that the $\mathbb{Z}_2$ symmetry $\si_n^x \to -
\si_n^x$ is spontaneously broken and $\si_n^x$ develops long-range order.
Similar arguments as above show that this phase has one Majorana mode of
type $b$ at the left end and one of type $a$ at the right end.

\textbf{Phase C}\--- One of the roots lies inside the unit circle
while two lie outside.
This is the case if $|\mu| ~\gg ~|J_x|, ~|\nu|, ~|J_y|$. Eq. (\ref{ham2}) then
implies that the $\mathbb{Z}_2$ symmetry $\tau_{n+1/2}^x \to - \tau_{n+1/2}^x$ is
spontaneously broken and $\tau_{n+1/2}^x$ develops long-range order.
This phase has no end Majorana modes and is therefore non-topological.

\textbf{Phase D}\--- All three roots lie outside the unit circle.
This is the case if
$|J_y| ~\gg ~|J_x|, ~|\mu|, ~|\nu|$. Eq. (\ref{ham1}) then implies that the
$\mathbb{Z}_2$ symmetry $\si_n^y \to - \si_n^y$ is spontaneously broken and $\si_n^y$
develops long-range order. Here, the end mode structure is switched compared
with case (B) in that there is one Majorana mode of
type $a$ at the left end and one of type $b$ at the right end.

We remark that this model, which has a duality property in the spin
values, is not symmetric under an exchange of $a$ and $b$ Majorana modes;
the Hamiltonian in Eq. \eqref{ham3} lacks a term of the form $a_n b_{n-2}$.
As a result, there is no phase analogous to phase $A$ in which there are
two Majorana modes of type $a$ at the left end and two of type $b$ at the
right end of a chain.

We now discuss the stability of the phases under changes in the different
parameters. Note that all points $(\la_1,\la_2,\la_3)$ in region A can be
smoothly taken to each other without
any of the roots crossing the unit circle, subject to the restriction that at
least one of them is real and the complex roots come in conjugate pairs. (Note
that a root can be made to pass through zero from the negative real side to
the positive real side by taking $J_y$ through zero. Similarly, a root can be
made to pass through $\infty$ from the very large positive real side to the
very large negative real side by taking $\nu$ through zero. In either case,
the root does not pass through the unit circle). Since the three roots and
the four parameters (up to a common scale) are related to each other by Eq.
(\ref{inv}), we see that it is not necessary that a point in region A must
correspond to $\nu$ being much larger than the other three parameters. However,
any point in region A can be smoothly taken, without crossing a phase boundary,
to a region in which $\nu$ is much larger than the other three parameters.

The unit circle $|\la| = 1$ corresponds to real values of $k$ in $\la =
e^{ik}$, i.e., the energy vanishes at some real momentum $k$ lying in
the range $[0,2\pi]$. So the
condition that none of the roots lie on the unit circle means that the
energy is gapped away from zero. The fact that all points in region A
are smoothly connected to each other without crossing the unit circle
means that they are all in the same phase, i.e., $\tau_{n+1/2}^y$ has
long-range order. On the other hand, to go from one phase to another (say,
from A to B), at least one of the roots must go through the unit circle,
so that the energy must at some point touch zero for some real momentum.

To sum up, we have shown that the model at hand has four distinct phases
characterized by spin ordering of $\si^{x/y}$ or $\tau^{x/y}$ operators. In terms of
Majorana end mode structure, phase (A) is the most interesting in that it is the only one
distinct from those found in the simple Kitaev chain and it hosts two independent
Majorana modes at each end.

\subsection{Phase Diagram}

\begin{figure}[h]
\begin{center} \begin{tabular}{ccc}
\epsfig{figure=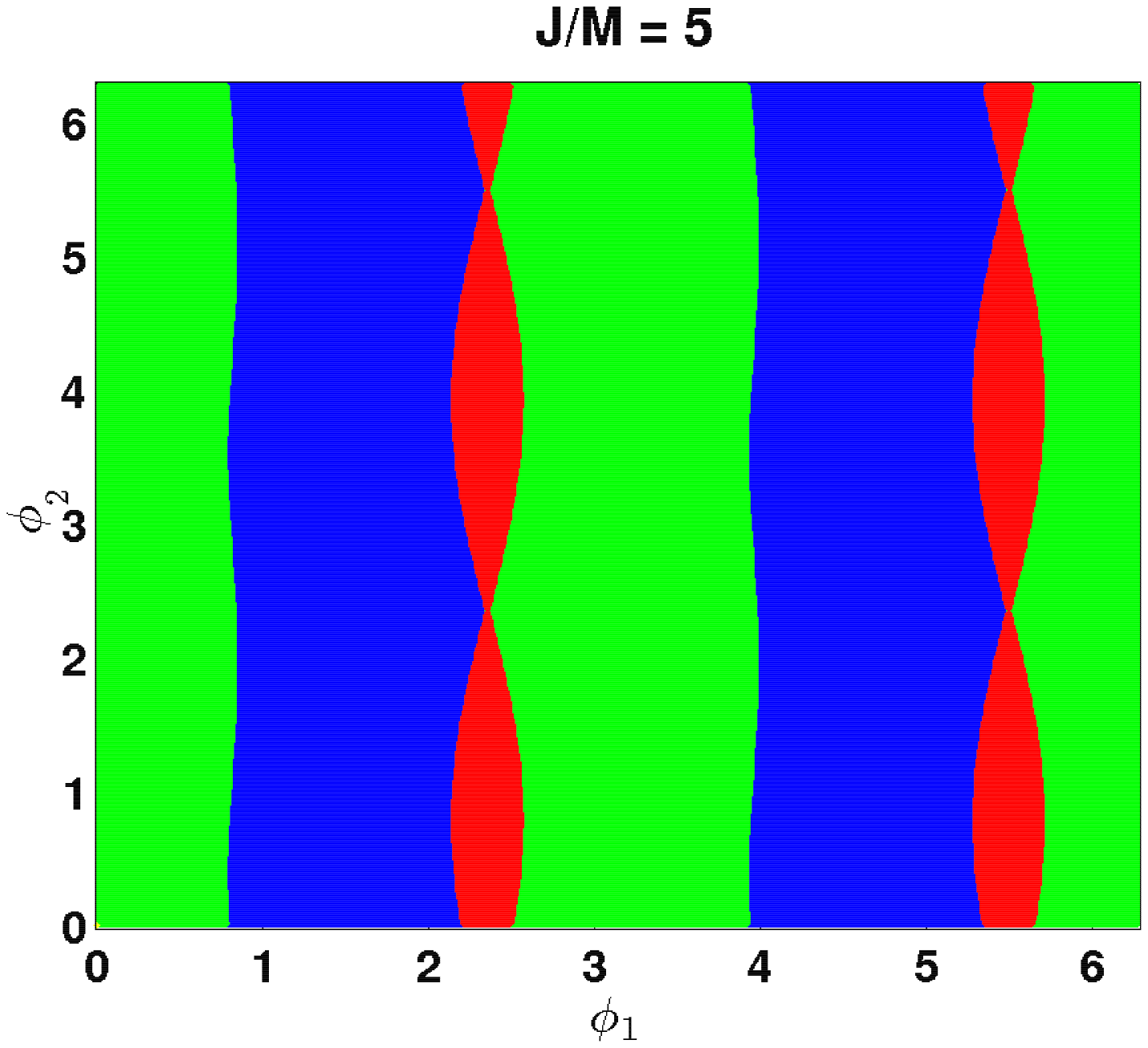,width=1.9in,height=1.87in,clip=} &
\epsfig{figure=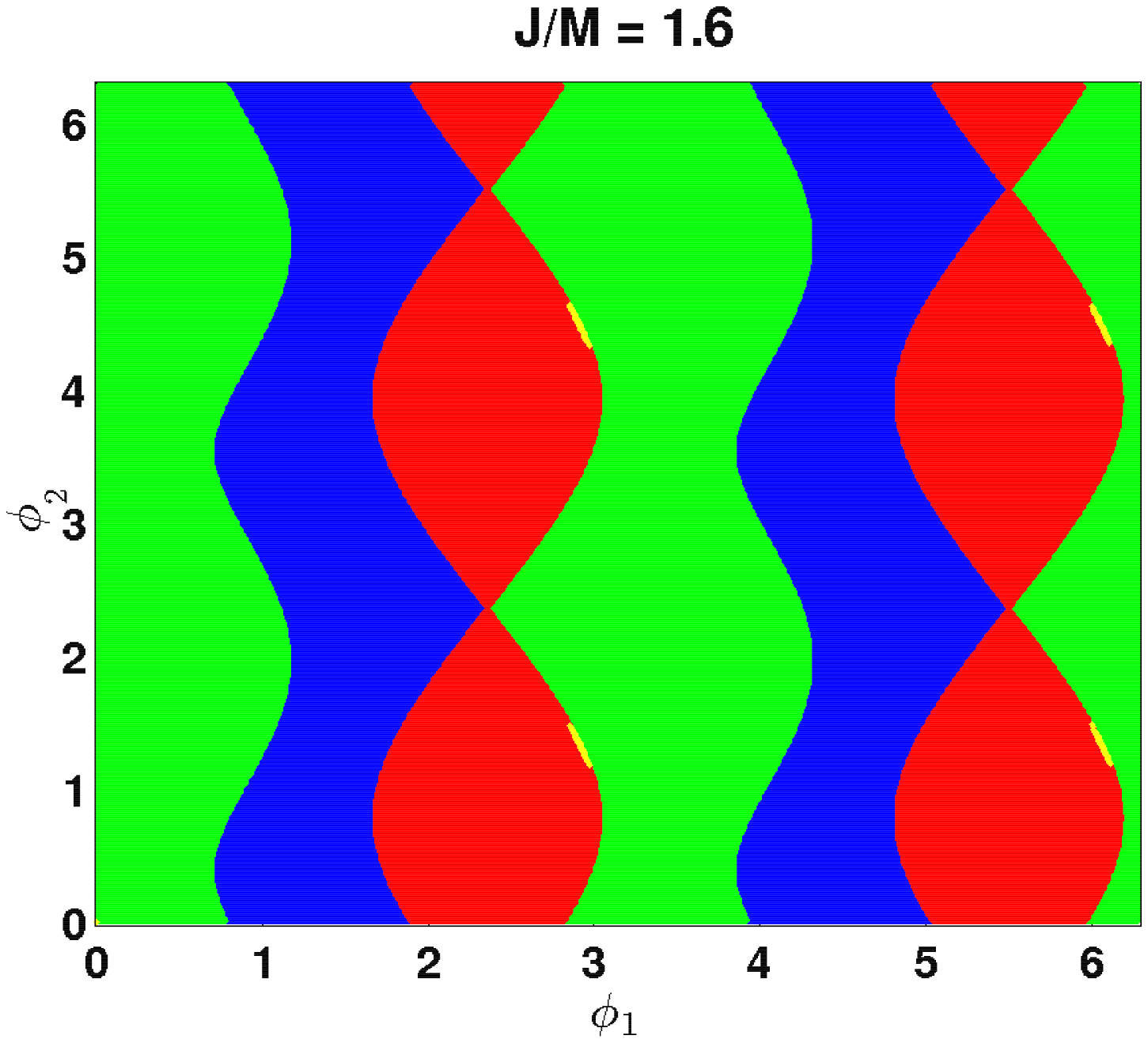,width=1.9in,height=1.87in,clip=} &
\epsfig{figure=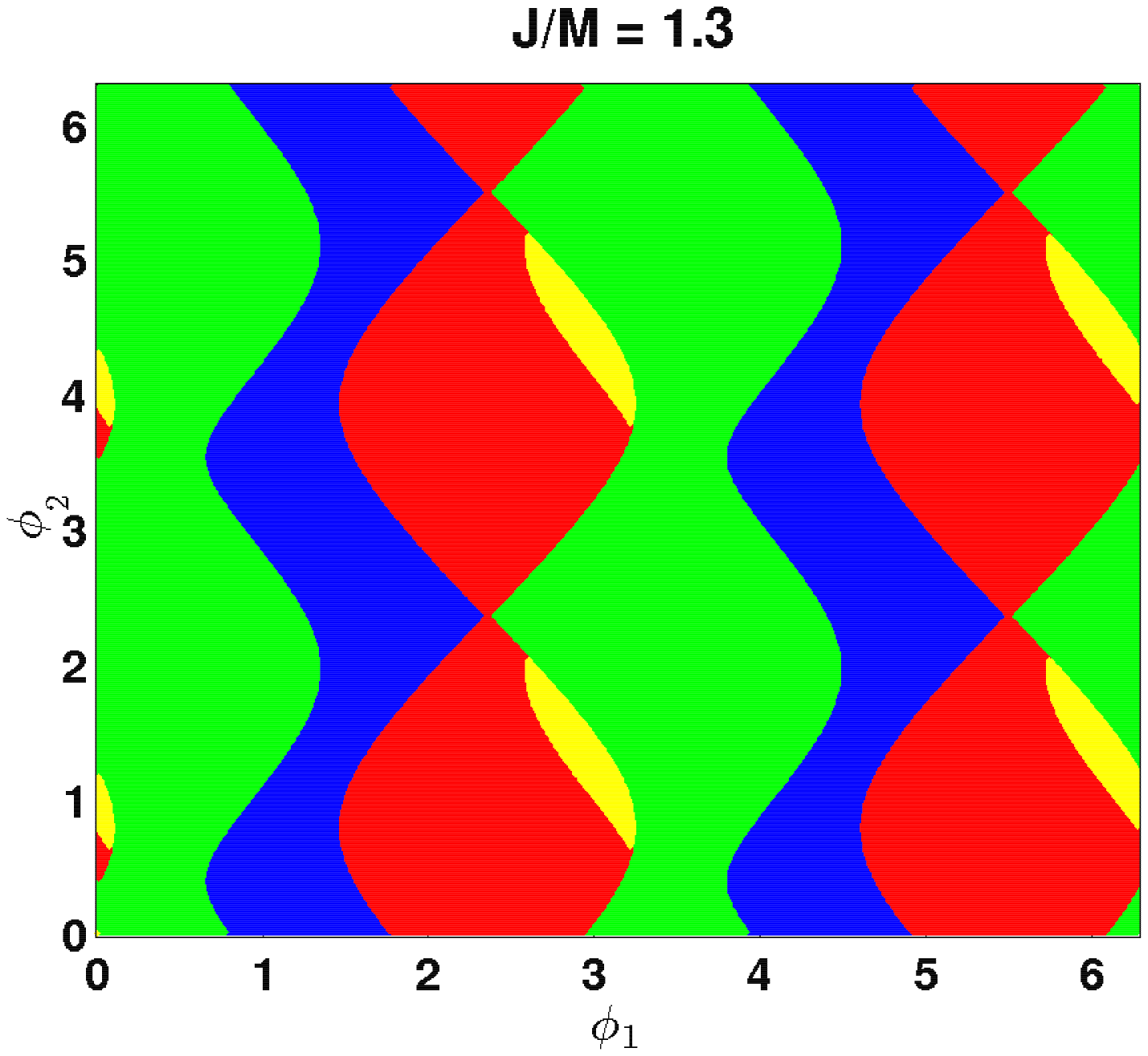,width=1.9in,height=1.87in,clip=} \\
\epsfig{figure=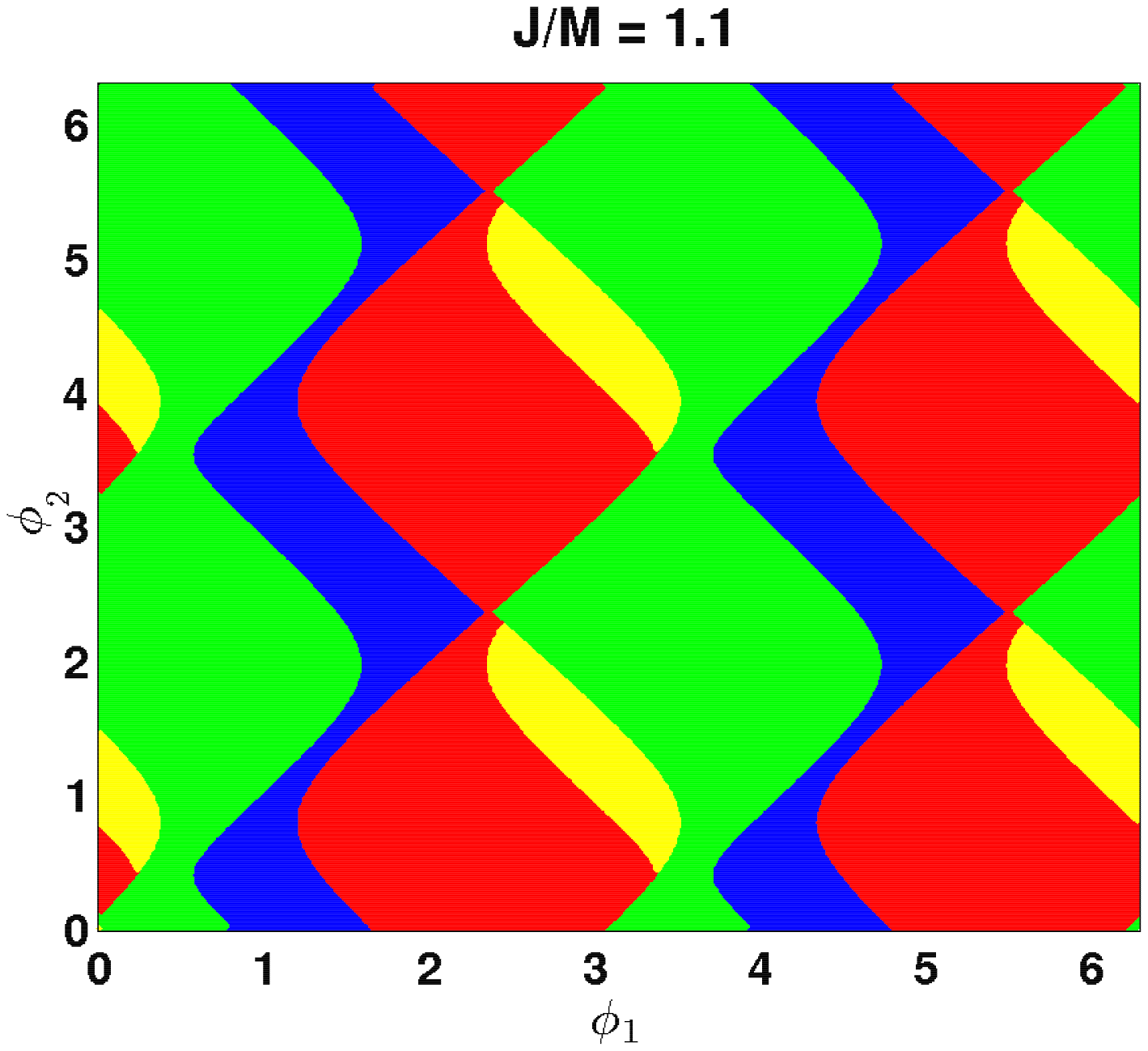,width=1.9in,height=1.87in,clip=} &
\epsfig{figure=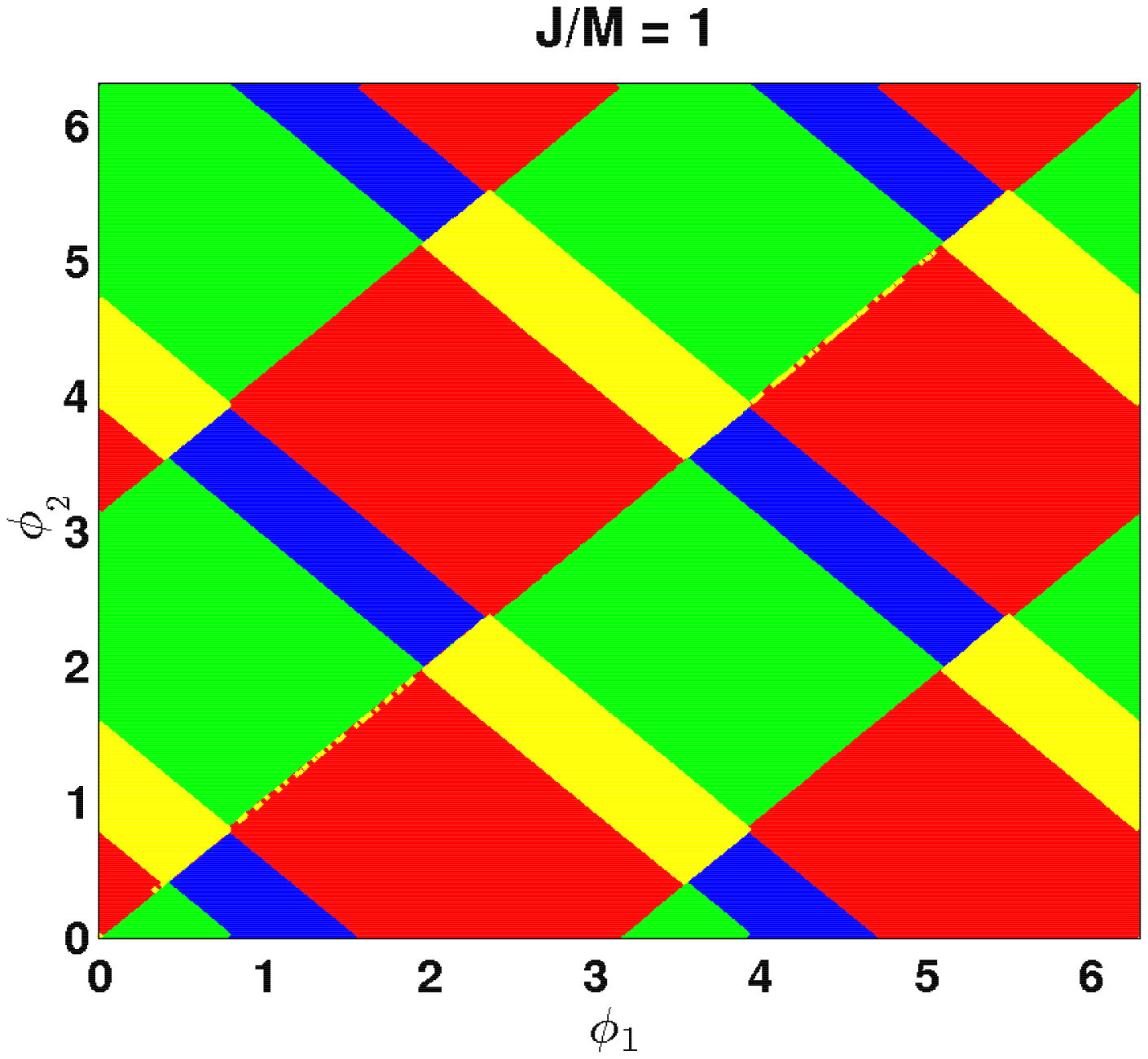,width=1.9in,height=1.87in,clip=} &
\epsfig{figure=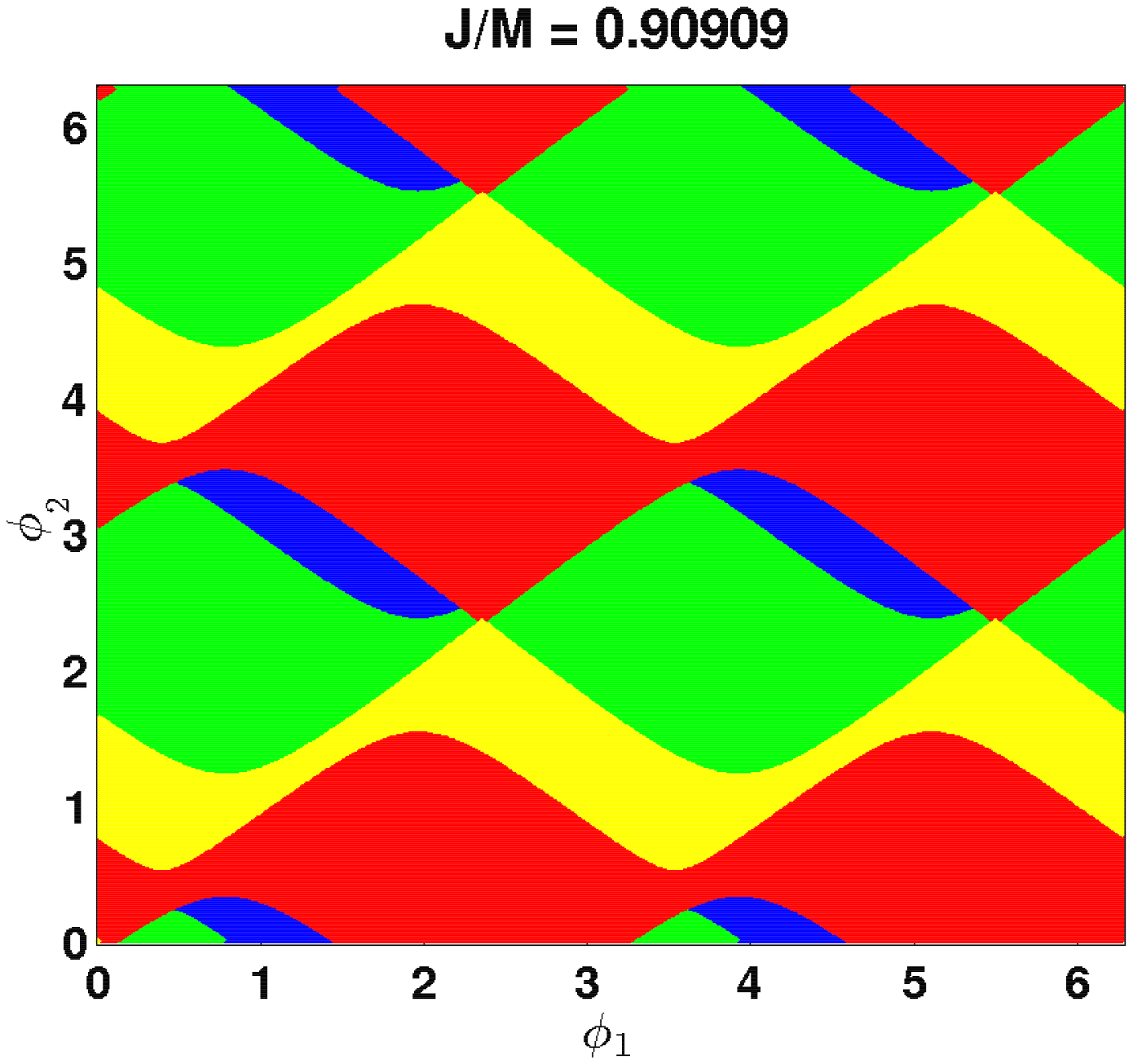,width=1.9in,height=1.87in,clip=} \\
\epsfig{figure=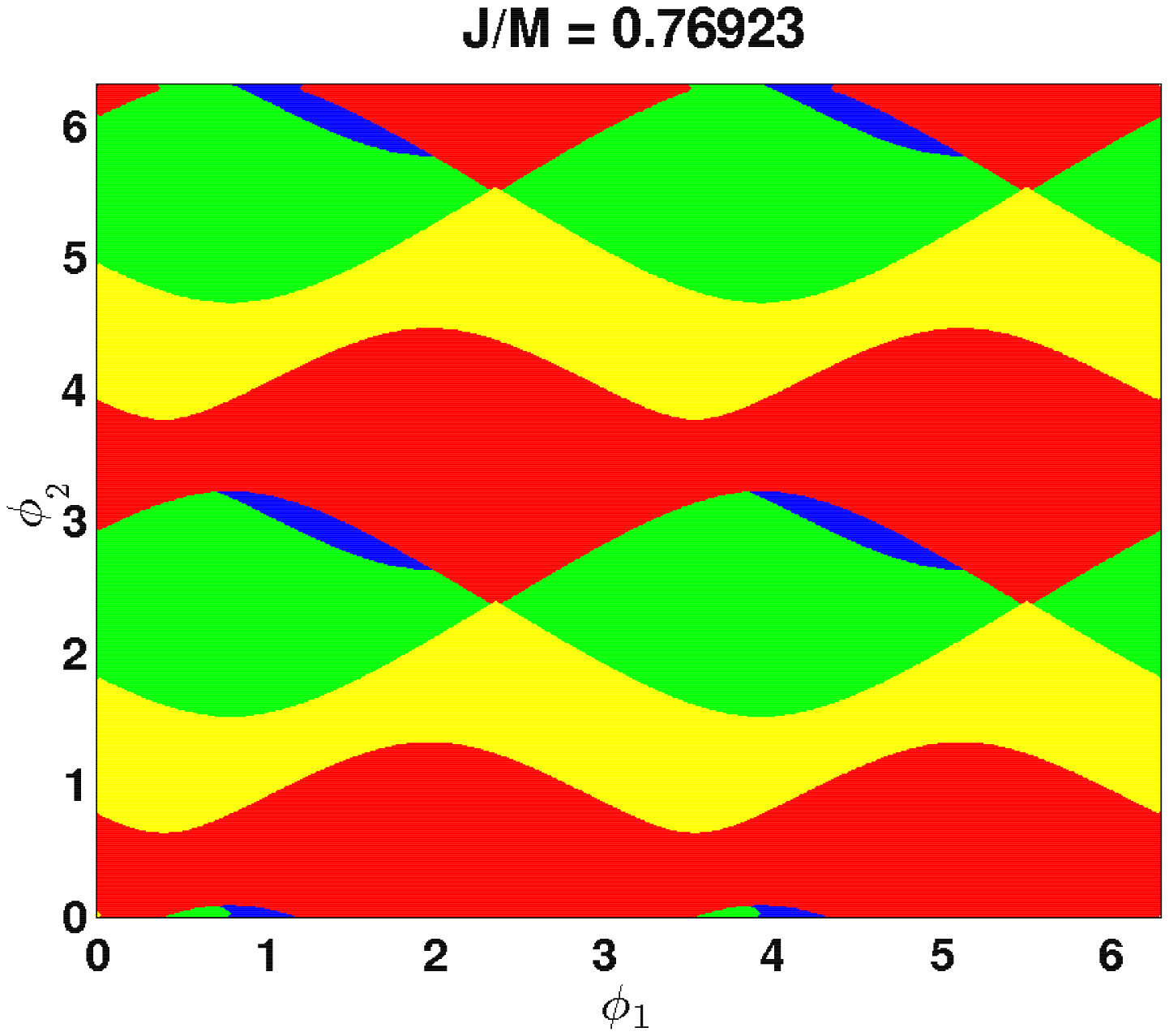,width=1.9in,height=1.87in,clip=} &
\epsfig{figure=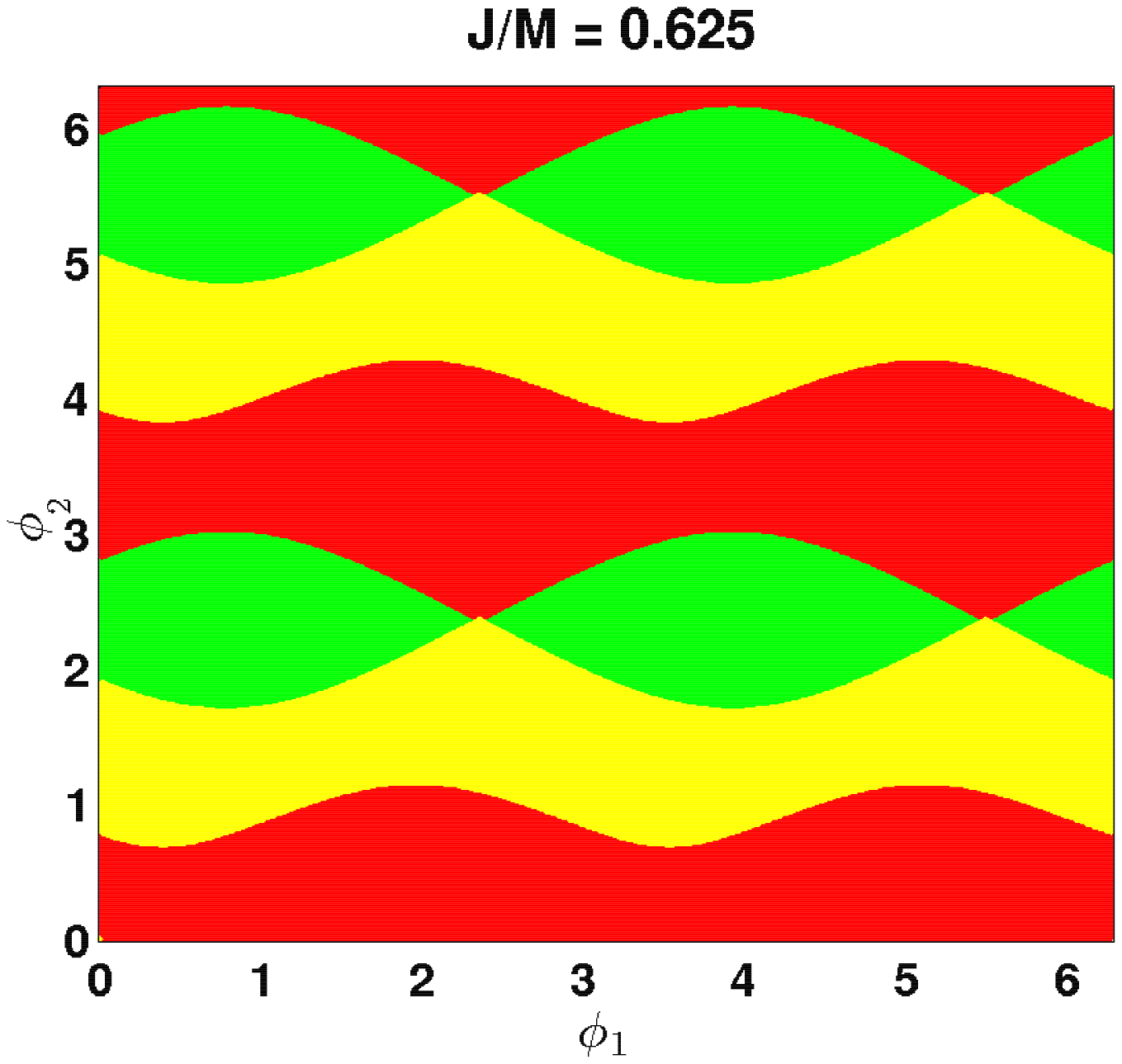,width=1.9in,height=1.87in,clip=} &
\epsfig{figure=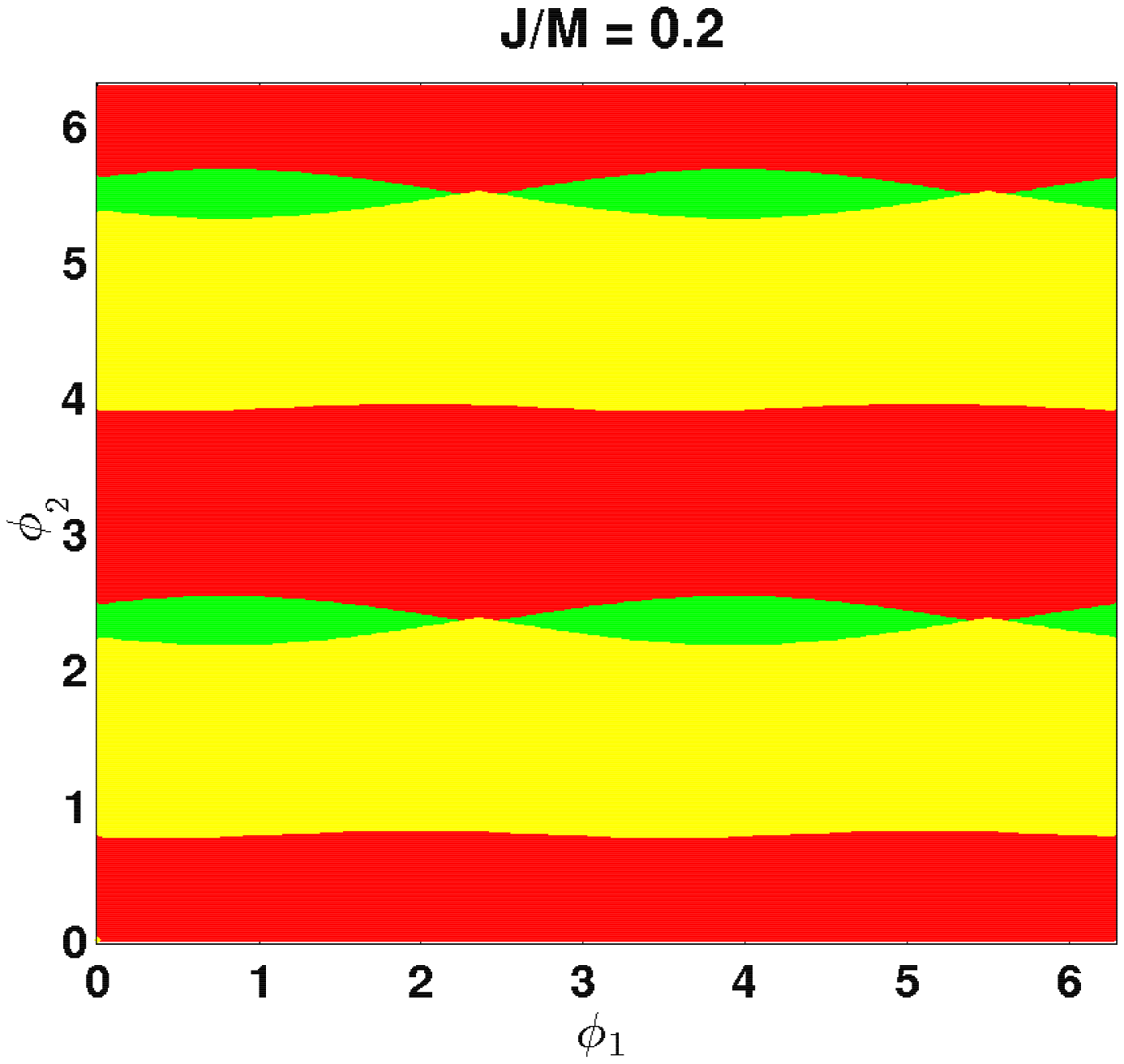,width=1.9in,height=1.87in,clip=} \\
& \epsfig{figure=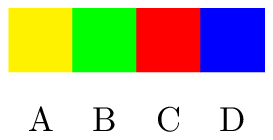,width=0.9in,height=0.5in,clip=} & \\
\end{tabular} \end{center}
\caption{Phase diagrams of Eq.~(\ref{ham1}) for $J/M ~= ~5, ~1.6, ~1.3, ~
1.1, ~1, ~1/1.1, ~1/1.3, ~1/1.6$ and $1/5$. The different colors represent
phases characterized by different number of Majorana modes at each end of a
finite system in the fermion language and long-range ordering of different
operators in the spin language. These phases correspond to A (yellow, two
end Majorana modes, $\tau^y$ ordering), B (green, one end Majorana mode,
$\si^x$ ordering), C (red, no end Majorana modes, $\tau^x$ ordering) and D
(blue, one end Majorana mode with roles of $a$ and $b$ interchanged compared
to phase B, $\si^y$ ordering). The panel at the bottom shows the colors of
the four phases which would be in increasing shades of darkness in a black
and white picture.} \label{fig:ising_phase} \end{figure}

Having identified the phases, we now study the phase diagram for different
values in parameter space. Since the phase does not change if
all the four parameters are scaled by the same number, we only have
a three-dimensional parameter space to consider. For convenience, we
recast the four couplings in terms of new parameters:
\bea J_x &=& J ~\cos \phi_1, ~~~~~~J_y ~=~ J ~\sin \phi_1, \non \\
\mu &=& M ~\cos \phi_2, ~~~~~~\nu ~=~ M ~\sin \phi_2, \label{jmphi} \eea
enabling us to study the phases as functions of $\phi_1$, $\phi_2$ and $J/M$.
We present the resultant phase diagrams in Fig.~\ref{fig:ising_phase}
where the nine panels correspond to different values of $J/M$.
 In each figure, the $x$ and $y$ axis correspond
respectively to $\phi_1$ and $\phi_2$ lying in the range $0$ to $2\pi$. Note
that every figure is invariant under $\phi_1 \to \phi_1 + \pi$ or $\phi_2 \to
\phi_2 + \pi$ or both. This is because these transformations either do not
change any of the roots or change the signs of all the roots, as we can see
from Eq. (\ref{root}) and both situations leave the phase unchanged.

We see from Fig.~\ref{fig:ising_phase} that only phases B (green) and D
(yellow) survive for $J/M \to \infty$, and only phases A (blue) and C (red)
survive for $J/M \to 0$. There is a
critical value of $J/M = (1+\sqrt{5})/2 \simeq 1.618$ above which phase A is
completely absent. This is given by the fact that if the three roots
$\la_1$, $\la_2$ and $\la_3$ all lie on or within the unit circle, then
one can show that
\beq \frac{J}{M} ~=~ \sqrt{ \frac{(\la_1 + \la_2 + \la_3)^2 + (\la_1 \la_2
\la_3)^2}{1 + (\la_1 \la_2 + \la_2 \la_3 + \la_3 \la_1)^2}} \eeq
has a maximum value of $(1+\sqrt{5})/2$ corresponding to $(\la_1, \la_2,
\la_3) = (1,1,-(3-\sqrt{5})/2)$ or $(-1,-1,(3-\sqrt{5})/2)$. Similarly, there
is a critical value of $J/M = (\sqrt{5}-1)/2 \simeq 0.618$ below which phase
D is completely absent. Hence the golden ratio determines the critical value
of $J/M$ for hosting more than three phases.

The most complex figure corresponds to $J/M = 1$. We see that the
phase boundary lines correspond to straight lines crossing each other at
$90^o$. This can be understood analytically as follows. On a phase boundary,
we have $\la = e^{ik}$, with real $k$, in Eq. (\ref{root}). If $k \ne 0$ or
$\pi$, we can take the real and imaginary parts of the equation to obtain
\bea (\nu + J_y) ~\cos (3k/2) ~+~ (\mu + J_x) ~\cos (k/2) &=& 0, \non \\
(\nu - J_y) ~\sin (3k/2) ~-~ (\mu - J_x) ~\sin (k/2) &=& 0. \eea
Using the identities $\cos (3k/2) / \cos (k/2) = 2 \cos k -1$ and
$\sin (3k/2) / \sin (k/2) = 2 \cos k +1$, we obtain
%\bea \cos k & = & \nu - J_y + J_x - \mu}{2(\nu - J_y)} \non \\
%&=& \frac{\mu + J_x - J_y - \nu}{2(\nu + J_y)}. \eea
$\cos k = (\nu + J_y - J_x - \mu)/(2(\nu + J_y)) = (\mu - J_x + J_y -
\nu)/(2(\nu - J_y))$.
Eliminating $\cos k$ from the above equations, we find
$J_x J_y - J_y^2 ~=~ \mu \nu - \nu^2$. Using Eq. (\ref{jmphi}), we get
%\beq J ~[ \cos (2\phi_1) + \sin (2\phi_1) - 1] ~=~ M ~[ \cos (2\phi_2) +
%\sin (2\phi_2) - 1]. \eeq
$J ~[ \cos (2\phi_1) + \sin (2\phi_1) - 1] = M ~[ \cos (2\phi_2) +
\sin (2\phi_2) - 1]$
For $J/M=1$, this gives the relationship
\beq \sin (2\phi_1 + \pi/4) ~=~ \sin (2\phi_2 + \pi/4). \eeq
This implies that either $\phi_2 = \phi_1 + n \pi$ or $\phi_2 + \phi_1 = n \phi
+ \pi/4$, where $n$ is an integer. These describe some of the straight line
phase boundaries that one
sees in the figure for $J/M = 1$. Some other straight lines in that figure come
from the conditions that a single root lies at $\la = \pm 1$ (corresponding to
$k=0$ or $\pi$); these give the conditions $\nu + J_x + \mu + J_y = 0$ and
$\nu - J_x + \mu - J_y = 0$ respectively which, for $J/M =1$, give $\sin
(\phi_1 + \pi/4)= \pm \sin (\phi_2 + \pi/4)$. These give some additional
straight lines corresponding to $\phi_2 + \phi_1 = 2 n \pi \pm \pi/2$.

We have thus provided an explicit model of long-ranging hopping which
has rich spin features and phase diagrams. The model supports
phases that go beyond those of the Kitaev chain in their multiple
Majorana structure.

\section{Broken Time-reversal Symmetry: ~Complex Hopping and $\De$}

In this section, we consider a different generalization of our model,
one in which TRS is broken. As discussed in Sec.~II B, this
 model belongs to a different symmetry class from that of the Kitaev chain,
namely class D. We find that the fate of the zero
energy modes, topology and Majorana structure is unusual in that TRS breaking
yields a finite regime in the phase diagram that is gapless.

\subsection{Model and Phases}
\label{sec:NoTRSModel}
Our model is described by the Hamiltonian
\beq H ~=~ \sum_n ~\Big[ - w f_n^\dag f_{n+1} - w^* f_{n+1}^\dag f_n
+ \De f_n f_{n+1} + \De^* f_{n+1}^\dag f_n^\dag - \mu (f_n^\dag f_n -1/2)
\Big] , \label{ham6} \eeq
where $w$ and $\De$ may be complex. We now observe that $\De$ can be made
real by a phase transformation of $f_n$ and $f_n^\dag$, namely, changing
$f_n \to f_n e^{i\theta}$ and $f_n^\dag \to f_n^\dag e^{-i\theta}$ changes
the phase of $\De$ by $e^{i2\theta}$ without changing the phase of $w$. On the
other hand, the phase of $w$ cannot be changed by any transformation without
affecting the phase of $\De$. We can therefore assume without loss of
generality that $\De$ is real. Writing $w=w_0 e^{i\phi}$, where $w_0$ is
real and positive, we obtain
\beq H ~=~ \sum_n ~\Big[ - w_0 e^{i\phi} f_n^\dag f_{n+1} - w_0
e^{-i\phi} f_{n+1}^\dag f_n + \De (f_n f_{n+1} + f_{n+1}^\dag
f_n^\dag) - \mu (f_n^\dag f_n -1/2) \Big]. \label{ham7} \eeq
By making appropriate phase transformations, we can ensure that $\phi$ lies
in the range $[0,\pi/2]$; we therefore study the properties of this
model only within this range. If we use the Jordan-Wigner transformation
in Eq.~\eqref{jw}, we find that Eq.~\eqref{ham7} takes the following form
in terms of a spin-1/2 chain
\beq H ~=~ -~ \frac{1}{2} \sum_n \Big[ (w_0 \cos \phi - \De) \si_n^x
\si_{n+1}^x ~+~ (w_0 \cos \phi + \De) \si_n^y \si_{n+1}^y ~+~ w_0 \sin
\phi ~(\si_n^x \si_{n+1}^y - \si_n^y \si_{n+1}^x) ~+~ \mu \si_n^z \Bigl]. \eeq

In momentum space, we find that Eq.~(\ref{ham7})
takes the form given in the first equation in Eq.~(\ref{ham5}), where
\beq h_k ~=~ (2w_0 \sin \phi \sin k) ~I ~-~ (2 w_0 \cos \phi \cos k ~+~ \mu)~
s^z ~+~ (2 \De \sin k) ~s^y. \label{hk2} \eeq
(Note that we have explicit TRS, $h^*_{-k} = h_k$, only if
$\phi = 0$ or $\pi$). The energy-momentum dispersion is given by
\beq E_k ~=~ 2w_0 \sin \phi \sin k ~\pm~ \sqrt{(2 w_0 \cos \phi
\cos k ~+~ \mu)^2 ~+~ 4 \De^2 \sin^2 k}. \eeq

\begin{figure}[h]
\begin{center} \begin{tabular}{ccc}
\epsfig{figure=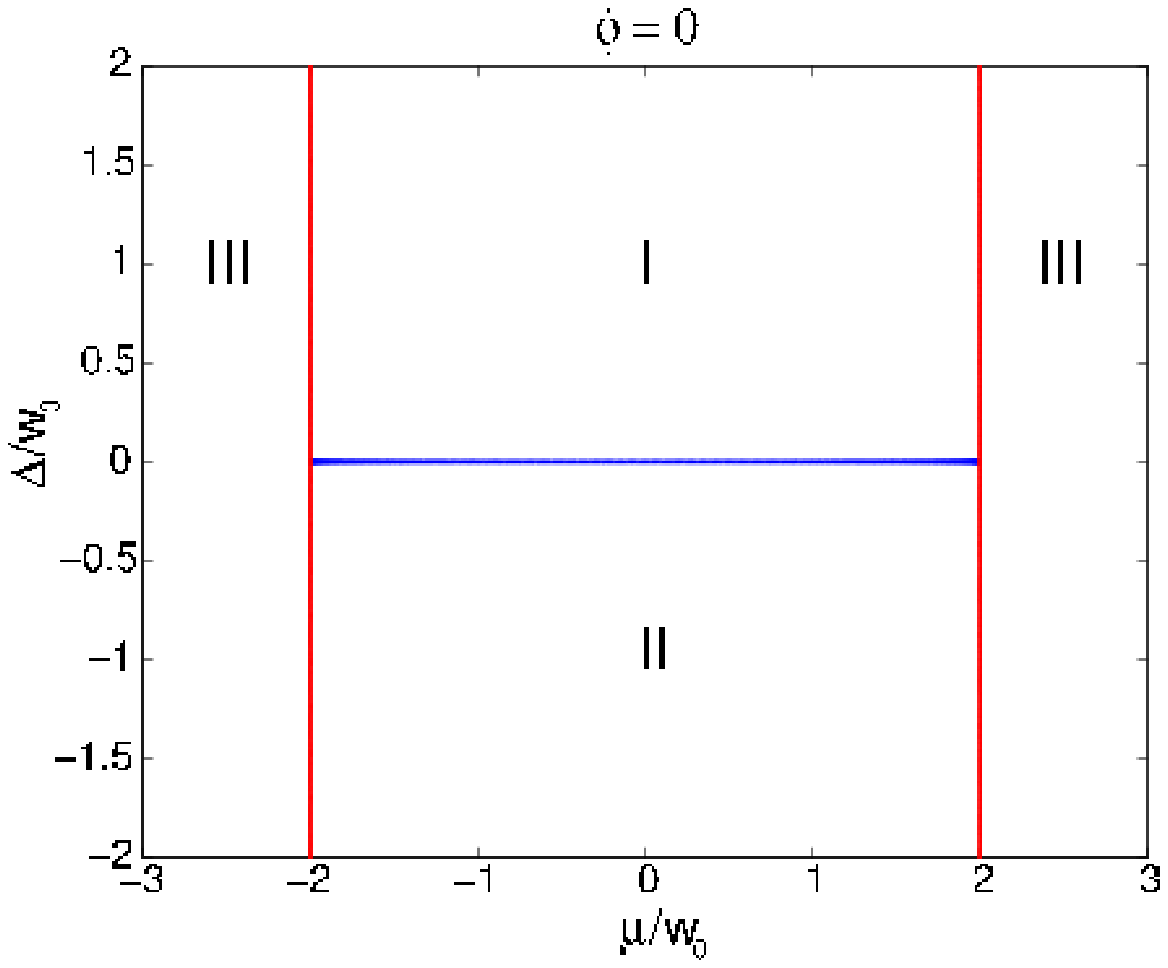,width=2.3in,height=1.8in,clip=} &
\epsfig{figure=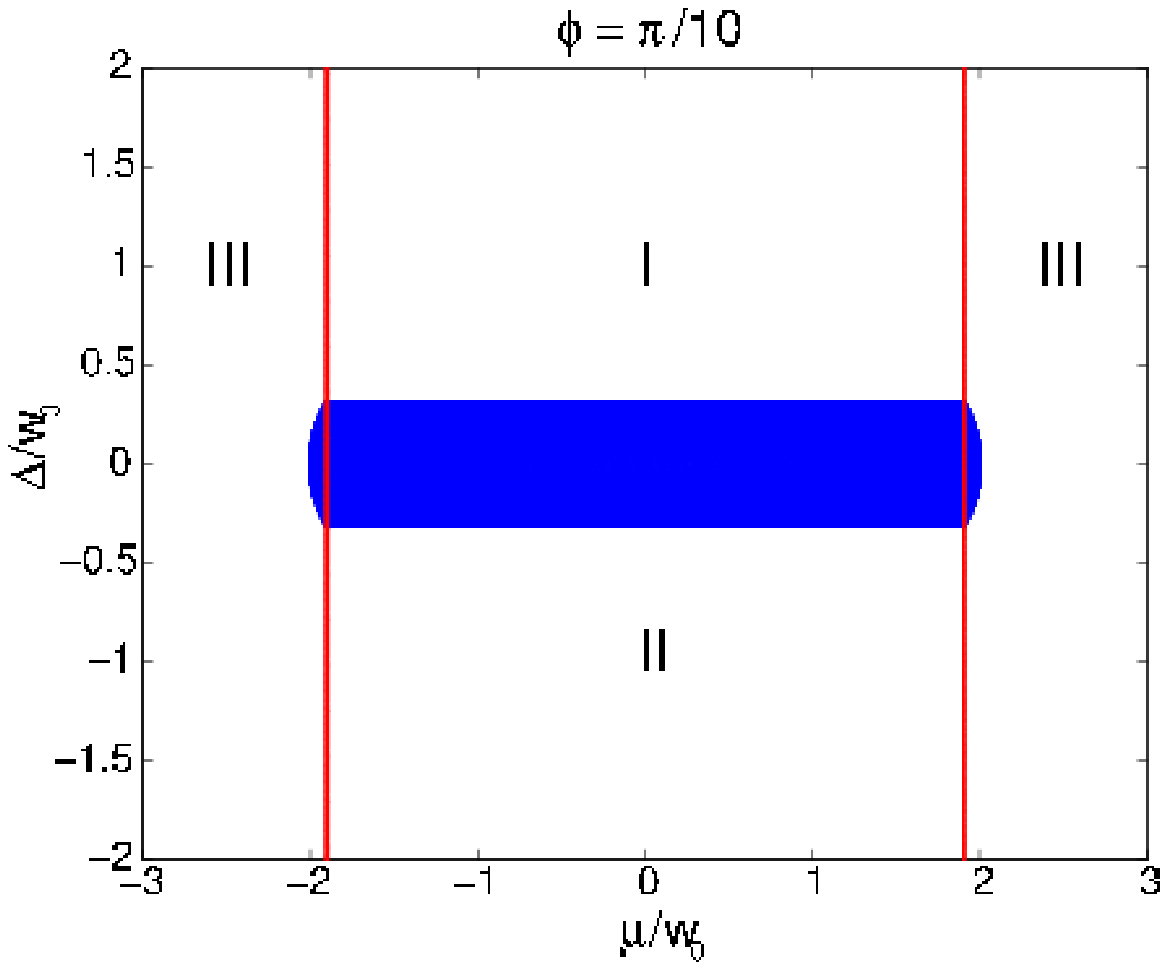,width=2.3in,height=1.8in,clip=} &
\epsfig{figure=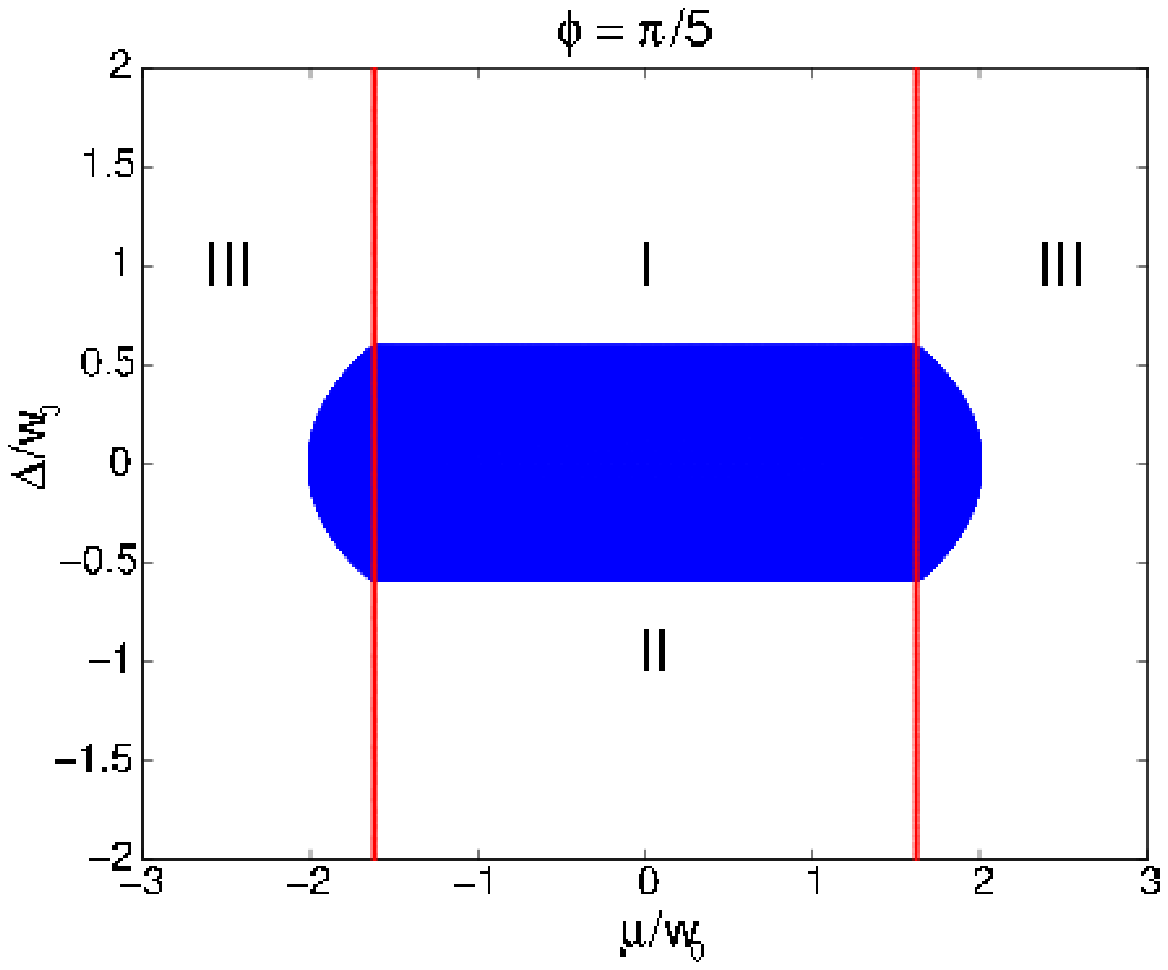,width=2.3in,height=1.8in,clip=} \\
\epsfig{figure=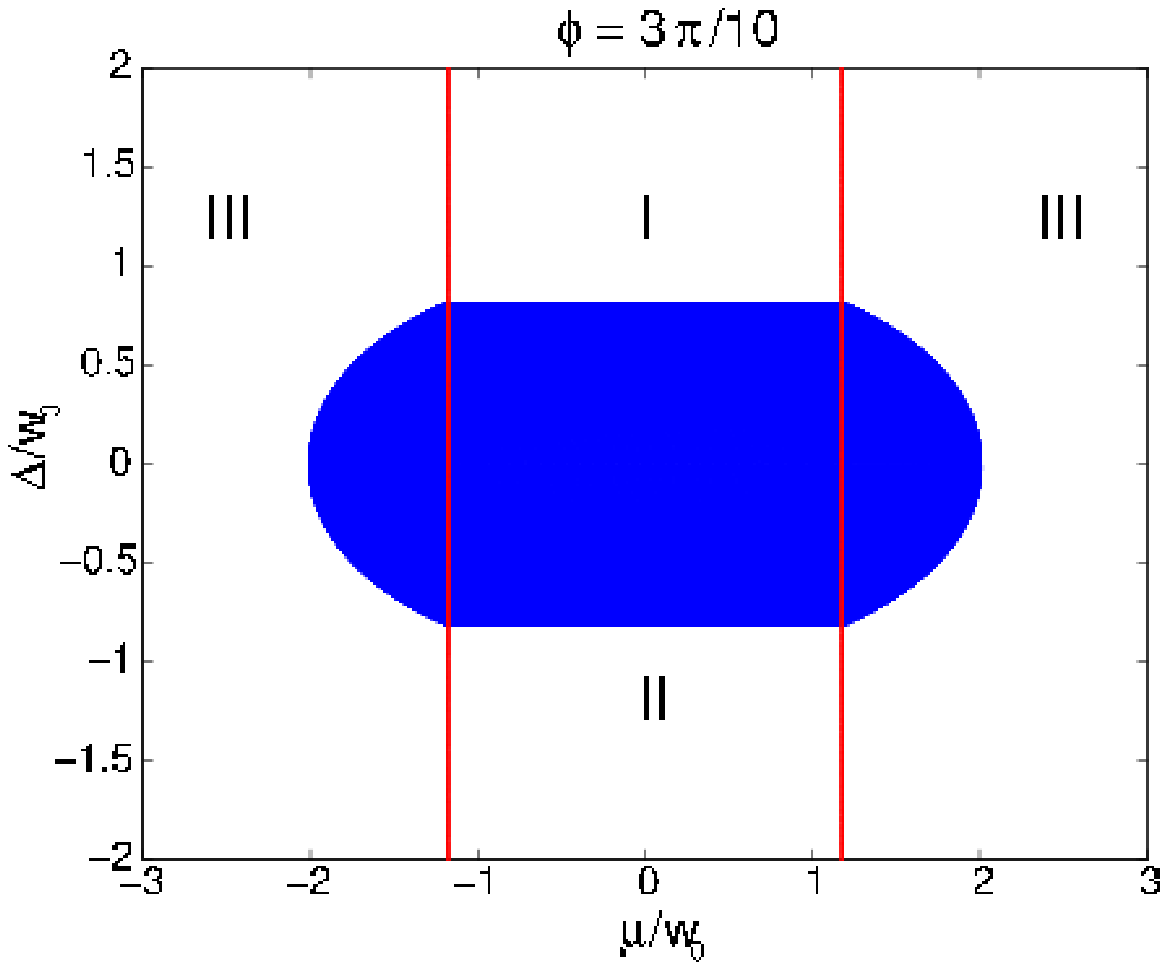,width=2.3in,height=1.8in,clip=} &
\epsfig{figure=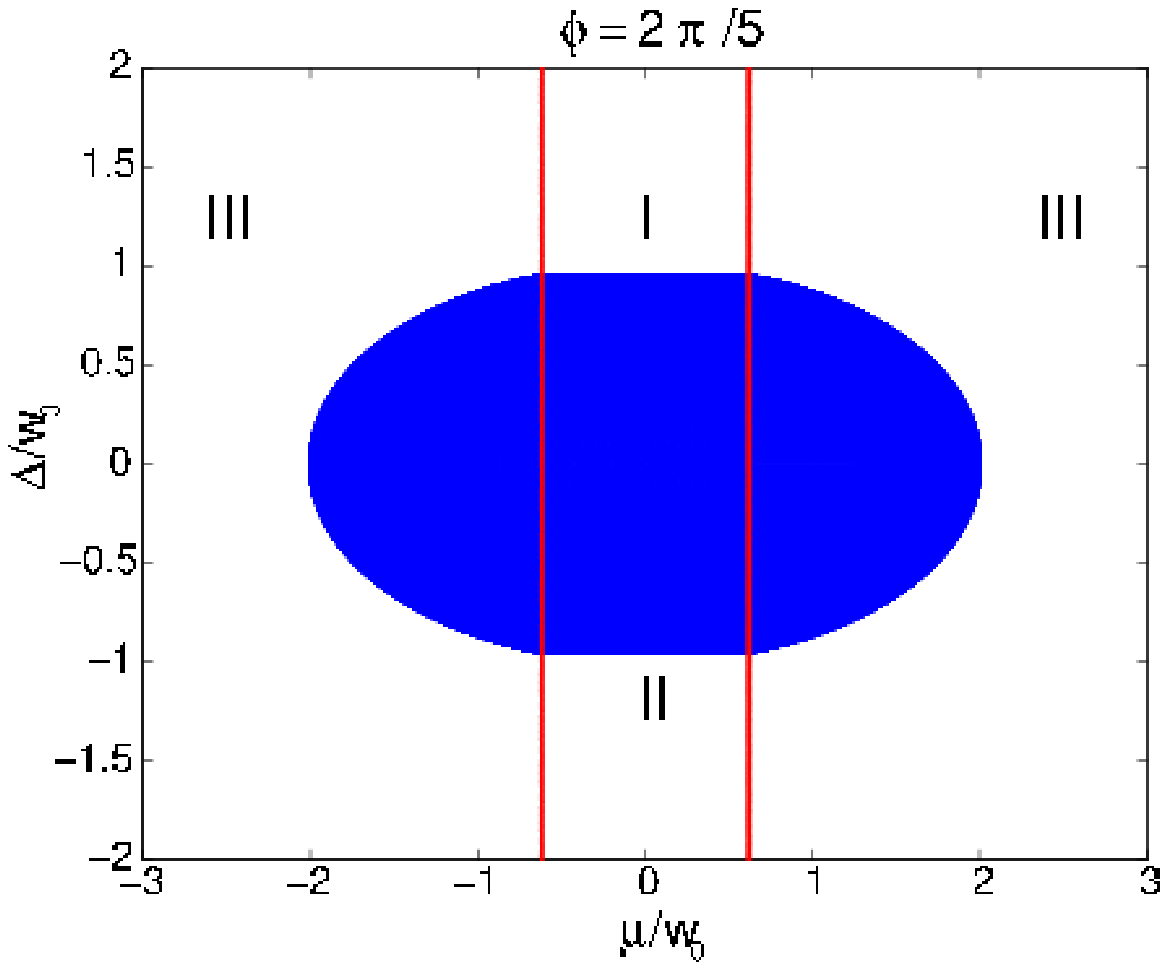,width=2.3in,height=1.8in,clip=} &
\epsfig{figure=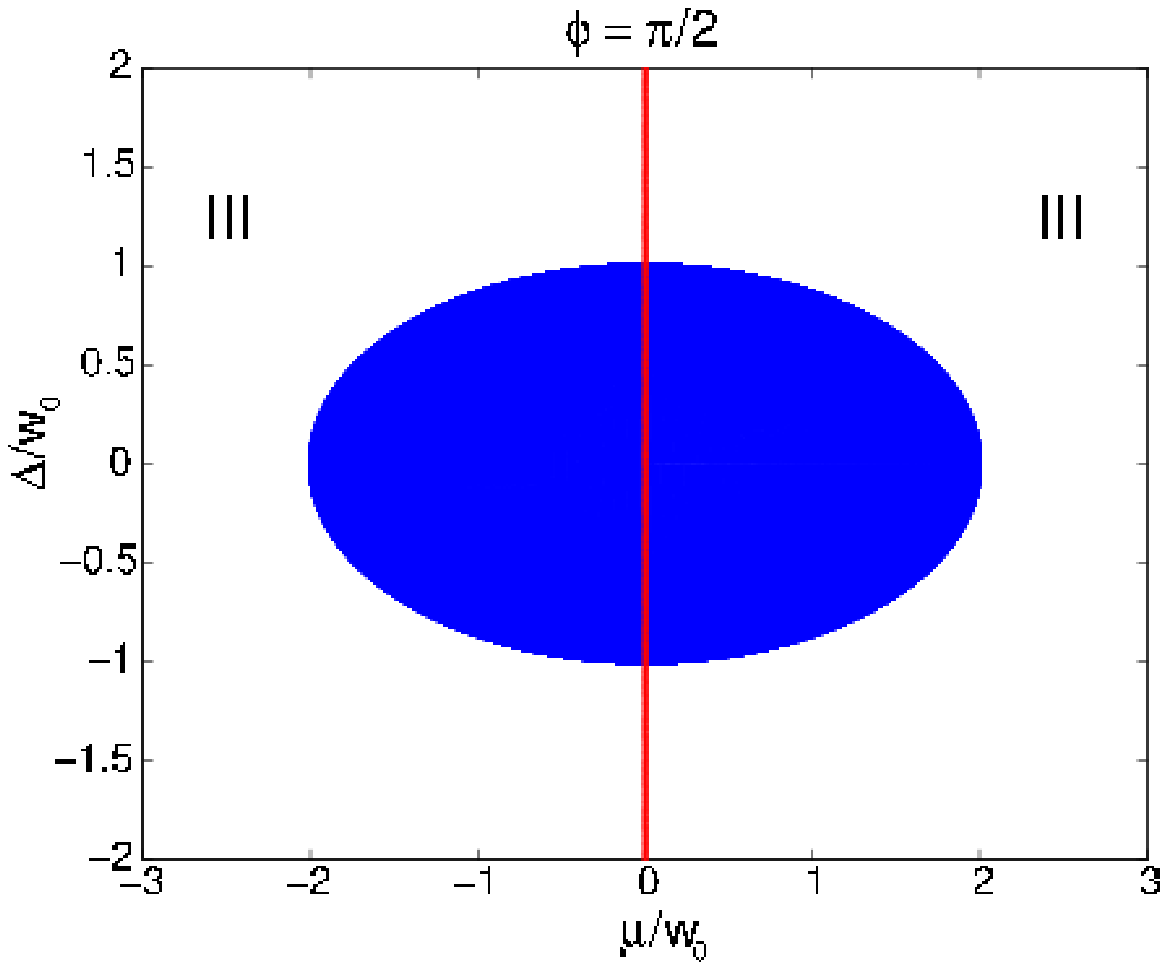,width=2.3in,height=1.8in,clip=} \\
\end{tabular} \end{center}
\caption{Phase diagrams of Eq.~(\ref{ham7}) as a function of $\mu /w_0$ and
$\De /w_0$, for $\phi ~=~ 0, ~\pi/10, ~\pi/5, ~3\pi/10, ~2\pi/5$ and
$\pi/2$. The system is gapless on the vertical red lines and everywhere in
the blue shaded regions. The figures show the phases I and II which are
topological and phase III which is non-topological; the three phases exist
in all the figures except for the bottom right figure ($\phi = \pi/2$)
where phases I and II do not exist.} \label{fig:tr_broken_phase} \end{figure}

The quantum critical points (or lines) are given by the condition
$E_k = 0$ for some value of $k$ lying in the range $[0,\pi]$, namely,
\beq (2 w_0 \cos \phi \cos k ~+~ \mu)^2 ~+~ 4 \De^2 \sin^2 k ~=~ 4 w_0^2
\sin^2 \phi \sin^2 k \label{qucrline} \eeq
for some $k$. To understand the complete phase diagram, we look at the quantum
critical points for six values of $\phi$ in the range $[0,\pi/2]$. The
resultant phase diagrams are shown in Fig.~\ref{fig:tr_broken_phase}. For all
values of $\phi$ (except for $0$ and $\pi/2$), the energy vanishes on two
lines and within a two-dimensional region. The two quantum critical lines are
given by the vertical lines $\mu/w_0 = \pm 2 \cos \phi$ (shown in red); on
these lines, the energy vanishes at $k=0$ or $\pi$. The two-dimensional
region (shown in blue) consists of a rectangle whose left and right sides
are capped by elliptical regions. The rectangle is bounded by the horizontal
lines given by $\De /w_0 = \pm \sin \phi$ and the vertical lines $\mu/w_0 =
\pm 2 \cos \phi$. Finally, the elliptical regions on the sides of the rectangle
always touch the points $(\mu/w_0, \De/w_0) = (\pm 2, 0)$.

So, to describe the evolution from the TRS preserved case
($\phi=0$) to the full broken case ($\phi=\pi/2$), at $\phi=0$, we have the phase diagram
respected by the Kitaev chain (the first figure in Fig.~\ref{fig:tr_broken_phase}).
Topological phases exist vertically below and above the gapless line running from $(\mu/w_0,
\De/w_0) = (-2, 0)$ to $(2,0)$ while the region $|\mu/w_0|>2$ are non-topological. As
seen in Fig.~\ref{fig:tr_broken_phase}, as $\phi$
 deviates from zero, the vertical phase boundaries ($\mu/w_0 = \pm 2 \cos
\phi$) approach one another,
the gapless line along $\De=0$ grows to a two-dimensional gapless regime that consists of a rectangular
region between the vertical phase boundaries and two elliptical caps on the sides. By invoking
adiabaticity, and as shown rigorously below, we find that the topological regions persist
above and below the gapless regions, and are bound between the vertical phase boundaries while
the rest of the gapped regime remains non-topological. Finally, upon
reaching $\phi = \pi/2$ (the last figure in
Fig.~\ref{fig:tr_broken_phase}), the two vertical lines merge into a single
line given by $\mu = 0$, shrinking the topological region to non-existence and the
two-dimensional region forms a single
ellipse given by $\mu^2 + 4 \De^2 = 4 w_0^2$.

The gapless regions, shown in blue in Fig.~\ref{fig:tr_broken_phase}, are
quite unusual. For each value of $k$ (except 0 and $\pi$) and $\phi$,
Eq.~\eqref{qucrline} describes an ellipse in the plane of $\mu/w_0$ and
$\De/w_0$. The gapless regions arise when all the ellipses for different
values of $k$ are combined. This is reminiscent of
the Kitaev model of spin-1/2's on a hexagonal lattice; that model also has
a gapless region in which the energy vanishes at different points in the
two-dimensional Brillouin zone~\cite{kitaev2}.

Turning to the fate of zero energy end modes, consider Eq.~(\ref{ham7})
expressed in terms of Majorana fermions:
\beq H ~=~ - ~\frac{i}{2} ~\sum_n ~\Big[ w_0 \cos \phi ~(a_n b_{n+1} ~+~
a_n b_{n-1}) ~-~ \De (a_n b_{n+1} - a_n b_{n-1}) ~+~ \mu ~a_n b_n ~+~ w_0
\sin \phi ~(a_n a_{n+1} ~+~ b_n b_{n+1}) \Big]. \label{ham8} \eeq
(Based on the time-reversal transformation of $a_n$ and $b_n$ discussed in
Sec. IV B, we again see that that this Hamiltonian does not have TRS
unless $\phi = 0$ or $\pi$).
The Heisenberg equations of motion following from Eq.~(\ref{ham8}) give
\bea w_0 \cos \phi ~(b_{n+1} ~+~ b_{n-1}) ~-~ \De ~(b_{n+1} ~-~ b_{n-1}) ~+~
\mu ~b_n ~+~ w_0 \sin \phi ~(a_{n+1} - a_{n-1}) &=& 0, \non \\
w_0 \cos \phi ~(a_{n+1} ~+~ a_{n-1}) ~-~ \De ~(a_{n-1} ~-~ a_{n+1}) ~+~
\mu ~a_n ~+~ w_0 \sin \phi ~(b_{n-1} - b_{n+1}) &=& 0, \label{eom3} \eea
for the zero energy modes. Eqs.~(\ref{eom3}) show that the zero energy modes
couple $a_n$ and $b_n$ if $\sin \phi \ne 0$, i.e., if TRS
is broken. In terms of Fig.~\ref{fig:tr_broken_phase}, we find that the
the region lying between the two vertical lines but excluding the rectangular
region forms a topological phase. In this phase, a long chain has one zero
energy Majorana mode at each end; these modes have real wave functions
(involving both $a_n$ and $b_n$), and they are separated from all the other
modes by a finite energy gap. We have shown this using numerical calculations
but we can also understand it analytically as follows.

Consider a general quadratic Majorana Hamiltonian of the form
\beq H ~=~ i \sum_{m,n=1}^{2N} ~c_m M_{mn} c_n, \label{hammaj} \eeq
where $M$ is a real antisymmetric matrix; hence $iM$ is Hermitian.
(For instance, this is the Hamiltonian we get for a $N$-site system if we
define the operators $c_n$ in terms of $a_n$ and $b_n$ as $c_{2n-1} = a_n$
and $c_{2n} = b_n$ for $n=1,2,\cdots,N$). One can show that the
non-zero eigenvalues of $iM$ come in pairs
$\pm \la_j$ (where $\la_j > 0$), and the corresponding eigenvectors are
complex conjugates of each other, $x_j$ and $x^*_j$; this is because
$iM x_j = \la_j x_j$ implies $iM x^*_j = - \la_j x^*_j$.
The number of zero eigenvalues of $iM$ must be even, and one can choose
those eigenvectors to be real: if $iM x_j = 0$, we have $iM x^*_j = 0$,
and we can then obtain real eigenvectors by taking the linear combinations
$x_j + x^*_j$ and $i(x_j - x^*_j)$.

Now, let us consider a Hamiltonian with TRS which has, say,
$q$ zero energy Majorana modes at the left end of a long chain. Let us assume
that the bulk modes are gapped at zero energy, i.e., there are no bulk states
with energies lying in the range $[-E_0,E_0]$, where $E_0$ is a positive
quantity. Next, let us add a small TRS breaking perturbation
to the Hamiltonian. By adiabaticity, the $q$ zero energy modes at the left end
 continues to lie within the bulk gap, but they need not remain at zero
energy. However, we know from the arguments in the previous paragraph, that
they can only move away from zero energy in pairs. Hence, if $q$ is odd,
one Majorana mode must remain at zero energy with a real eigenvector.
Since we know from Sec. II A that the case with $\phi = 0$, which
has TRS, phases I and II
are topological and have one Majorana mode at each end of a chain, this must
continue to remain true if we make $\phi$ non-zero, as long as the bulk
gap does not close. However, the nature of the end Majorana mode changes
from the case with TRS (where it involves only $a_n$ or only $b_n$) to
the case with broken TRS (where it involves both $a_n$ and $b_n$).

\subsection{General Quadratic Majorana Hamiltonian and Topological Phases}

As seen above, TRS breaking results in couplings of the
$a$ fermions amongst themselves, and similarly couplings amongst the $b$
fermions. In this section, we will consider the most general Hamiltonian
which is quadratic in terms of the Majorana fermions $a_n$ and $b_n$, thus
generalizing the arguments presented in Sec. IV A
(on long-range hopping) to include TRS breaking.
This enables us to discuss the $\mathbb{Z}_2$-valued invariant which appears
in such a system.

For an infinitely long chain, such a Hamiltonian can be written as
\beq H ~=~ -i \sum_{r=-\infty}^\infty~ \sum_{n=-\infty}^\infty ~J_r ~a_n
b_{n+r} ~-i ~\sum_{r=1}^\infty~ \sum_{n=-\infty}^\infty ~[K_r ~a_n a_{n+r} ~+~
L_r ~b_n b_{n+r}], \label{ham9} \eeq
where $J_r, ~K_r, L_r$ are all real parameters. In momentum space, this can
be written as in the first equation in Eq.~(\ref{ham5}), where
\bea h_k &=& -2 \sum_{r=-\infty}^\infty ~[J_r ~\cos (kr) ~s^z ~+~ J_r \sin
(kr) ~ s^y] \non \\
&& + ~2 ~\sum_{r=1}^\infty ~[(K_r + L_r) ~\sin (kr) ~I ~+~ (K_r - L_r) ~\sin
(kr) ~s^x]. \label{hk3} \eea
The energy-momentum dispersion is given by
\bea E_k &=& 2 ~\sum_{r=1}^\infty ~[(K_r + L_r) ~\sin (kr) \non \\
&& \pm~ 2 ~\sqrt{\left( \sum_{r=-\infty}^\infty ~J_r ~\cos (kr)
\right)^2 ~+~ \left( \sum_{r=-\infty}^\infty ~J_r \sin (kr) \right)^2 ~+~
\left( \sum_{r=1}^\infty (K_r - L_r) ~\sin (kr) \right)^2}. \eea

We now observe that Eq.~(\ref{hk3}) has TRS, i.e.,
$h^*_{-k} = h_k$ for all $k$, only if $K_r = L_r = 0$ for all $r$. In that
case, as indicated in Secs. IV A-B, $h_k$ defines a vector in
the $y-z$ plane as in Eq.~(\ref{vk}); since this plane contains the origin,
we have already seen that we can define a winding number around the origin
as in Eq.~(\ref{wind}), and this is a TI taking values in
$\mathbb{Z}$ (the set of all
integers). However, if some of the $K_r$ or $L_r$ are non-zero, then $h_k$
does not have TRS and has all four components (along $I, ~s^x, ~
s^y$ and $s^z$) in general; one cannot define
a winding number around the origin for a closed curve (generated by $k$
going from 0 to $2\pi$) if the curve is in more than two dimensions.
On the other hand, at $k=0$ and $\pi$, $h_k$ only has a component along $s^z$
given by $h(0) = -2 \sum_r J_r$ and $h(\pi) = - 2 \sum_r (-1)^r J_r$
respectively. Thus, for the gapped regions in parameter space,
we explicitly see that we can employ the bulk TI
defined in Eq.~\eqref{eq:bulkZ2top}, yielding
\beq \nu_{bulk} = \mbox{sgn} \left( 4 \sum_r J_r \sum_q (-1)^q J_q \right). \eeq
This invariant takes the values $-1$ and 1 in the topological and
non-topological phases respectively; its value cannot change unless either
$h(0)$ or $h(\pi)$ crosses zero in which case the energy $E_0$ or $E_\pi$
must vanish. As an example, for the simple system
with complex hopping discussed in the previous subsection, Sec. \ref{sec:NoTRSModel},
we see that $\nu_{bulk} = \mbox{sgn} (\mu^2 - 4 w_0^2 \cos^2 \phi)$ is equal to $-1$ in
the regions $|\mu| < 2 w_0 \cos \phi$. Once we consider the fact that
in this regime, the system is gapped only for $|\De| > 2 w_0 \sin \phi$,
the condition $\nu_{bulk} = -1$ captures the correct topological regions
shown in Fig.~\ref{fig:tr_broken_phase}.

To explicitly analyze the nature of the end Majorana modes, we have
seen that in a system with TRS (with $K_r = L_r = 0$) there
can be any number of such modes at each end. For instance,
Fig.~\ref{fig:endmodes} (e)
shows a system in which only $J_2$ is non-zero and there are two zero
energy modes at each end of an open chain shown by $a_3$, $a_4$, $b_5$ and
$b_6$. However, if we now break TRS slightly by turning
on small values of $K_1$ and $L_1$ in the Hamiltonian, i.e., introduce terms
like $-i K_1 a_3 a_4 - i L_1 b_5 b_6$, the energies of these modes shift
away from zero to $\pm K_1$ and $\pm L_1$, destroying any zero
energy Majorana modes. One can see an odd-even effect manifest in that
if there are an odd number of zero energy Majorana modes at one end
of the chain in the presence of TRS, breaking the
symmetry would in general couple these modes. But given that there are
an odd number of states and that the Hamiltonian only permits states to come
in $\pm E$ pairs, unlike in the even case, one $E=0$ Majorana end mode has
to survive.

To conclude, a system with TRS can have an arbitrary number of
zero energy Majorana modes at each end of a long chain. If TRS is
then broken, pairs of these end modes move away from zero
energy. In general, therefore, we are left with either no Majorana modes
or only one Majorana mode depending on whether the parent system
with TRS has an even or an odd number of Majorana modes.

\section{Spatially Varying Potentials}

Having considered the effects of long-range hopping and TRS breaking on the
topological features of the Kitaev chain, we now treat the effect of
subjecting the chain to a spatially varying potential landscape. For the
discretized lattice, the potential landscape can be described in terms of a
site-dependent chemical potential. The last term containing the
global chemical potential $\mu$ in Eq.~\eqref{ham0} can be replaced
by one having a local chemical potential $\mu_n$. The topology of such
systems has been studied before~\cite{adagideli,shivamoggi,sau,ladder,niu,
period,tezuka,lang,motrunich,brouwer1,lobos,cook,pedrocchi,degottardi}

The phase diagram in Fig.~\ref{fig:tr_broken_phase} (a) shows that in the
the spatially homogeneous case, topological phases exist in the Kitaev
chain even with arbitrarily small superconductivity present ($\De \neq 0$) and
persist until a chemical potential $\mu$ that is of the order of the bandwidth
$w$ is applied~\cite{kitaev1}. Here we shall see that this is a singular limit and that for
spatially varying potential these features no longer persist~\cite{brouwer1,brouwer2,brouwer3}.
In fact, in general we will see that for any finite potential, a minimum
amount of superconductivity is required to reach the topological phase,
though there remain interesting counterexamples to this general rule (see
Fig.~\ref{fig:quasi})~\cite{motrunich}.

In what follows, we analyze three different kinds of potential landscapes,
expanding and building on our recent short work, Ref.~\cite{degottardi}. As the
simplest example and a direct application of the transfer matrix techniques
presented in Sec. II B, we obtain the phase diagram for
periodic potentials having short periods commensurate with the lattice (see also ~\cite{ladder,niu,period}).
We then develop our transfer matrix further to enable an extensive
study of more complex potentials. In particular, we provide a mapping
between the normal state properties of the wire and its topology in the superconducting state. Our result is a more general form of the mappings presented in Refs.~\cite{sau,adagideli}.
However, this transformation becomes singular at $\De = 1$, and thus cannot by itself be used to obtain the full phase diagram (see Sec. VI. B). The line $\De = 1$ may be treated separately (this regime corresponds to the
random field Ising model). We have also developed a mapping between the topology between the gap function $\De$ and $1 / \De$. Taken together, these procedures allows us to obtain the topological phase diagram and the decay length of the end Majorana wave functions in terms of the normal state localization properties of the system. This mapping enables us to leverage the vast literature on the localization
properties of 1D systems to the question of the topology of these systems
in the presence of superconductivity.

Given the close connection between normal state properties and topology, we find that a great
deal of insight can be gained by briefly reviewing the band structure of simple periodic systems, which is
done in Sec. VI A. In Sec. VI B, the mapping between normal state properties and topology will be developed. Finally, Sec. VI C presents an application of these methods to ultra-long periodic and quasiperiodic potentials.

\subsection{Periodic Potentials and band structure}

Given that we will forge a connection between the topological and normal
state properties of inhomogeneous systems, it is helpful to review some
basic properties of the normal state band structure~\cite{ladder,degottardi}.
The problem of an electron hopping in a periodic potential is described
by Eq.~\eqref{ham0} where we set $\De = 0$ and take the chemical potential
$\mu_n$ to depend on $n$ in a periodic way. For a state with energy $E$,
we get the discrete form of the Schr\"odinger equation
\beq -w (f_{n+1} + f_{n-1}) - \mu_n f_n ~=~ E f_n, \eeq
This can be written in the transfer matrix form
\beq \left( \begin{array}{c}
f_{n+1} \\
f_n \end{array} \right) ~=~ \left( \begin{array}{cc}
- \frac{1}{w} (E + \mu_n) & -1 \\
1 & 0 \end{array} \right) \left( \begin{array}{c}
f_n \\
f_{n-1} \end{array} \right).
\label{eq:normalmatrix}
 \eeq
The energy spectrum consists of those values of $E$ which admit
plane wave states, $f_n \sim e^{\pm i k n}$. For a system with period $q$,
the appropriate object to consider is
\beq \mathcal{A} = \prod_{n = 1}^q \left( \begin{array}{cc}
- \frac{1}{w} (E + \mu_n) & -1 \\
1 & 0 \end{array} \right). \eeq
Since $\mathcal{A}$ is $2 \times 2$, its eigenvalues satisfy
\beq \la^2 - \left( \mbox{Tr} \mathcal{A} \right) \la +
\mbox{det} \mathcal{A} = 0. \eeq
Let $\la_\pm$ denote the two eigenvalues of $\mathcal{A}$.
Since $\mbox{det }\mathcal{A} = 1$, we have $\la_+ \la_- = 1$. Further,
$\la_+ + \la_-$ must be real. Hence
\beq \la_\pm =
\begin{cases}
e^{\pm i r} ~~~~~ \mbox{for} ~~|\mbox{Tr} ~\mathcal{A}| ~\le~ 2, \non \\
e^{\pm r} ~~~~~~ \mbox{for} ~~\mbox{Tr} ~\mathcal{A} ~>~ 2, \\
- e^{\pm r} ~~~~ \mbox{for} ~~\mbox{Tr} ~\mathcal{A} ~<~ -2, \\
\end{cases} \label{eq:eigen} \eeq
where $r$ is real. The object $\mbox{Tr}~\mathcal{A}$ is a $q^{th}$ order polynomial in $E$, and
the spectrum of the system corresponds to the those values of $E$ for which
$|\mbox{Tr} \mathcal{A}| \le 2$. A useful quantity which quantifies the
localization properties of the system is the Lyapunov exponent
$\gamma(W) \equiv \lim _{\begin{subarray}{l} \mathcal{N} \to
\infty \end{subarray}} \frac{1}{\mathcal{N}} \ln | \max (\la_\pm) |$. Fig. 5 presents
a detailed example of the normal state properties of a periodic system at zero energy as
well as anticipating the connection between the normal state and the topological properties developed
in Sec. IV B.

\begin{center}
\begin{figure}
\epsfig{figure=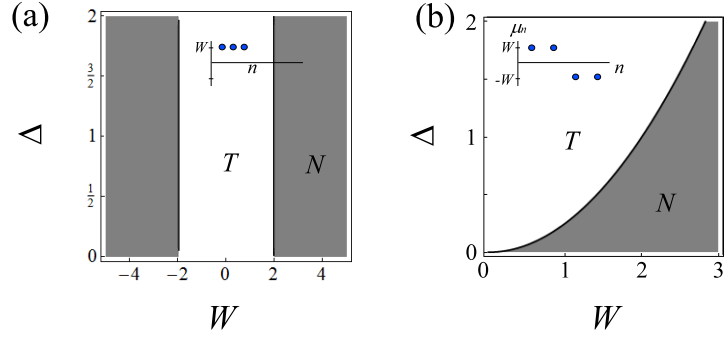,width=10cm}
\caption{Topological phase diagram showing the topological ($T$) and
non-topological ($N$) phases as a function of the potential strength $W$ and
the superconducting gap $\De$ for (a) a uniform potential $\mu_n = W$, and
(b) a periodic potential having the pattern $(W,W,-W,-W)$.}
\label{fig:periodic2} \end{figure} \end{center}

It is instructive to consider the case of simple periodic patterns as they
offer a great deal of insight into the properties of the phase diagrams in
more complicated situations. These features anticipate and provide
qualitative understanding of our central results in subsequent sections. For instance,
we will consider potentials of the form $\mu_n = W \epsilon_i$, where
$\epsilon_i$ is some inhomogeneous potential of fixed strength and zero mean.
A (very) naive caricature of such a system would be a period-2 potential of
the form $\mu_n =...,-W, W,...$ or a period-4 potential $\mu_n = ...,-W,-W,
W,W,...$. The topology of these simple periodic patterns may be obtained by taking the
transfer matrix $\mathcal{A} = \prod_{n = 1}^q \mathcal{A}_n$ introduced in Sec. II B, where
$q$ is the period of the potential. Given $\mathcal{A}$, TI $\nu$ may be straightforwardly found by applying the methods of Sec.~\ref{sec:boundaryinvariants} and Eq.~\eqref{eq:TIinvariant}, in particular. Such simple periodic patterns naturally arise in a spin ladder model in which a vortex degree of freedom controls the sign of the chemical potential in the corresponding fermion system~\cite{ladder}, a one-dimensional version of Kitaev's celebrated honeycomb model~\cite{kitaev2}.
Several simple examples are given in Fig.~\ref{fig:periodic2} and Table~\ref{tab:sectors}).

\begin{table}
\caption{Criteria for topological phases for a selection of periodic
potentials (we have set $w=1$).}
\label{tab:sectors}
\begin{center}
\begin{tabular}{|c|c|c|}
\hline
period & pattern & topological for \\
\hline
\hline
1 & $ \ldots,W,W, W, \ldots $ & $| W | < 2$ \\
\hline
2 & $ \ldots, W, - W,\ldots$ & $ \De > | W | / 2$ \\
\hline
4 & $ \ldots, W, W, W, -W,\ldots$ & $\De^2 > W^2/2 - 1 $ \\
4 & $ \ldots, W, W, -W, -W, \ldots $ & $ \De > W^2/4 $ \\
\hline
\end{tabular}
\end{center}
\end{table}

\begin{center}
\begin{figure}
\epsfig{figure=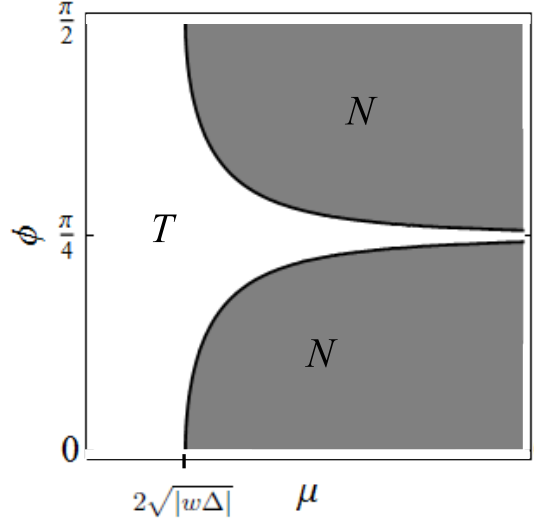,width=10cm}
\caption{Topological phase diagram for a period-4 potential, showing the
topological ($T$) and non-topological ($N$) phases as a function of $\mu$
and $\phi$.} \label{fig:period4}
\end{figure}
\end{center}

The presence of a phase offset in commensurate periodic potentials
can significantly affect the topological phase diagram.
As an illustrative example, let us consider the period-4 pattern given by
\beq \mu_n = \sqrt{2} \mu \cos \left( \frac{\pi }{4} (2 n + 1)
+ \phi \right). \eeq
Since shifting $\phi \to \phi + \pi/2$ is equivalent to shifting $n \to n+1$,
it is sufficient to study this problem in the range $0 \le \phi \le \pi/2$.
Upon finding the eigenvalues of the product of four transfer matrices
$A_4 A_3 A_2 A_1$ of the form given in Eq.~\eqref{an} below, we find that
the phase boundary between the topological and non-topological phases is given
by
\beq 4 \sqrt{2} | w \De | = \mu^2 \sqrt{1 + \cos (4 \phi)}.
\label{eq:period4} \eeq
The phase diagram is shown in Fig.~\ref{fig:period4}. The fact that the
boundary between the topological and non-topological phases diverges in $\mu$
as $\phi \to \pi /4$ can be understood as follows. For $\phi = 0$,
we know that the pattern of the chemical potential takes the form
$P = + + - -... $ and thus has the phase boundary given by $\mu = 2
\sqrt{|w \De|}$ (see the last line of Table I). For $\phi = \pi/4$, the
pattern assumes the form $Q = + 0 - 0 +...$. In the limit $\mu \to \infty$
(in which the hopping plays no role), the eigenstates of the system
are localized to each site. Given that $\mu = 0$ at every other site,
the system must contain bound Majorana modes and therefore must be
topological. The general shape of the phase boundary then follows from the
fact that it connects the points $(2 \sqrt{|w \De|},0)$ and
$(\infty,\pi/4)$ in the $(\mu, \phi)$ plane. In general we conjecture that
for any periodic pattern, $\mu_n = \mu \cos (\pi Q n + \phi)$ where $Q$ is
a rational number, if the chemical potential vanishes at certain sites for
some value of $\phi$, a `finger' of the topological phase will extend up to
$\mu = \infty$ for that value of $\phi$. As will be seen in the next
section, the sensitivity of the phase diagram to potentials vanishing
at certain sites is quite ubiquitous and results in a non-analytic
behavior of the phase boundary for disordered systems.

\subsection{Mapping to Normal Systems and Duality}
\label{sec:mapping}

As observed in Ref.~\cite{motrunich}, the product of such transfer matrices,
$\mathcal{A} \equiv \prod_{n = 1}^{\mathcal{N}} A_n$, is strongly reminiscent
of that used to determine localization properties of the normal state
Anderson disorder problem, i.e. Eq.~(\ref{eq:normalmatrix}). We build
on this observation by explicitly mapping the topological phase diagram to the
normal state properties of the system (similar to a procedure developed in~\cite{sau}). This mapping is performed in two steps.
The first allows us to determine topological regions in the superconducting
system in the range $0 < \De < 1$ based on a knowledge of its normal counterpart (same spatially varying potential but no superconductivity). The second is a duality which relates the topological phase diagram for a given superconducting gap $\De$ to that corresponding to strength $1/\De$. Taken together, these provide a complete mapping between the normal state and the topology of the superconducting system.

\textbf{Mapping to Normal Systems} -- In Sec. II B, we introduced
the transfer matrix relevant to a superconducting wire having spatially
varying potentials:
\beq A_n = \left( \begin{array}{cc}
-\frac{\mu_n}{\De + w} & \frac{\De - w}{\De + w} \\
1 & 0 \end{array} \right). \label{an} \eeq

For $0 < \De < 1$, we perform a similarity transformation
$A_n = \sqrt{\delta} S \tilde{A}_n S^{-1}$ with $S = \mbox{diag} (\delta^{1/4},
1/\delta^{1/4} )$ and $\delta = \frac{1-\De}{1+\De}$. The matrices
$\tilde{A}_n$ are of the form shown in Eq.~(\ref{eq:matrixA}) with $\De
\to 0$ and $\mu \to \mu_n/\sqrt{1-\De^2}$. This immediately gives
\beq \mathcal{A}(W,\De)=\left( \sqrt{ \frac{1-\De}{1+\De}}
\right)^\mathcal{N} S \mathcal{A}\left(W/\sqrt{1-\De^2},0\right) S^{-1}.
\label{eq:trans} \eeq
Taking the logarithm of the eigenvalues of Eq.~(\ref{eq:trans}),
the condition that $|\la_2| = 1$ is given by
\beq \gamma(W,\De) = \gamma \left( \frac{W}{\sqrt{1- \De^2}},0 \right) -
\frac{1}{2} \ln \left(\frac{1+\De}{1-\De} \right),
\label{eq:phaseboundary} \eeq
reminiscent of a result in~\cite{sau}, where we have defined the Lyapunov
exponent $\gamma(W,\De) \equiv \lim _{\begin{subarray}{l} \mathcal{N} \to
\infty \end{subarray}} \frac{1}{\mathcal{N}} \ln | \la_2 (W,\De) |$; the
Lyapunov exponent is the inverse of the localization length, $\gamma(W,\De)
= 1/\ell(W,\De)$. In the limit $\gamma(W,\De) \to 0$, the system is gapless
and Eq.~(\ref{eq:phaseboundary}) describes the phase boundary separating the
topological and non-topological regions of the phase diagram. This
relation quantifies the observation in~\cite{motrunich,brouwer2} that in
general a critical amount of superconductivity must be applied before the
system is driven into a topological phase. For the case in which the system
is metallic (i.e., $\gamma(W,0) = 0$), any non-zero $\De$ will give rise to
a topological phase.

If $\mu_n = 0$ for all $n$, we see from Eq.~\eqref{an} that all the $A_n$ are
identical and have both eigenvalues less than 1 in magnitude if $\De > 0$ (we
are assuming that $w > 0$). This implies that $\ga (W=0,\De) < 0$.
Thus the line $W=0$ and $\De > 0$ will always lie in the topological phase.

\textbf{Duality} -- The similarity transformation
(Eq.~\eqref{eq:trans}) is only valid in the range $0 < \De < 1$. The form of
the phase diagram for $\De > 1$ may be obtained by noting that
the transformation
\bea \mu_n \to \mu_n/\De, ~~\De \to 1/\De,~~\mbox{and}~~
P \to \tilde{P} \label{eq:duality} \\
\mbox{where}~~P \to \tilde{P}:~ \{\mu_n\} \to \{(-1)^n \mu_n \}, \non \eea
leaves the eigenvalues of $\mathcal{A}$ unchanged for $\mathcal{N}$ even.
Thus, if a point $(W_0,\De_0<1)$ lies on the phase boundary of $P$, then
$(W_0/\De_0,1/\De_0)$ lies on the phase boundary of $\tilde{P}$. This
duality strongly constrains the form of the phase boundary in the cases in which
the distribution is invariant under the transformation in
Eq.~(\ref{eq:duality}).

An interesting illustration of this duality is provided by the periodic
patterns in Table~\ref{tab:sectors}. For example, note that the uniform case
(period 1) and the period 2 case are dual to each other. Indeed, taking
$\De \to 1/\De$ and $W \to W / \De$ for the equation $W = 2$ (phase
boundary for period 1 case), we obtain the phase boundary $\De = W/2$,
appropriate for the period 2 case. Similarly, the two period 4 patterns are
self-dual. For instance, taking $\De^2 = W^2 - 1$, and applying $\De
\to 1/\De$ and $W \to W / \De$ gives $1/\De^2 = W^2 / \De^2 - 1$
which is equivalent. The same holds true for $\De = W^2/4$.

\textbf{Random-field quantum Ising Chain} \--- Finally, at the point $\De = 1$, the system maps to the quantum
Ising chain subject to a spatially varying transverse field. As can be seen in Sec. \ref{sec:spinmap}, the
Jordan-Winger transformation applies even when the local chemical potential
$\mu_n$ has a spatial dependence and appears as a spatially varying field along the $z$-direction. Along the line
$\De/w=1$, we have $J_y=0$, and thus for a random distribution of $\mu_n$,
the system corresponds to the 1D random transverse field
Ising model, which has been studied in great depth~\cite{fisher,shivamoggi}.
 The matrix $\mathcal{A}(W,1)$
has the eigenvalues $\frac{1}{2^\mathcal{N}} \prod_{n = 1}^\mathcal{N} \mu_n$
and 0. Eq.~(\ref{eq:TIinvariant}) reveals that the phase boundary passes
through the point for which
\beq \langle \ln | \mu_n | \rangle = \ln 2, \label{eq:specialpoint1} \eeq
where $\langle \ln | \mu_n | \rangle \equiv \lim_{\mathcal{N}\to \infty}
\frac{1}{\mathcal{N}} \sum_{n=1}^\mathcal{N} \ln | \mu_n |$.

\begin{center} \begin{figure} \epsfig{figure=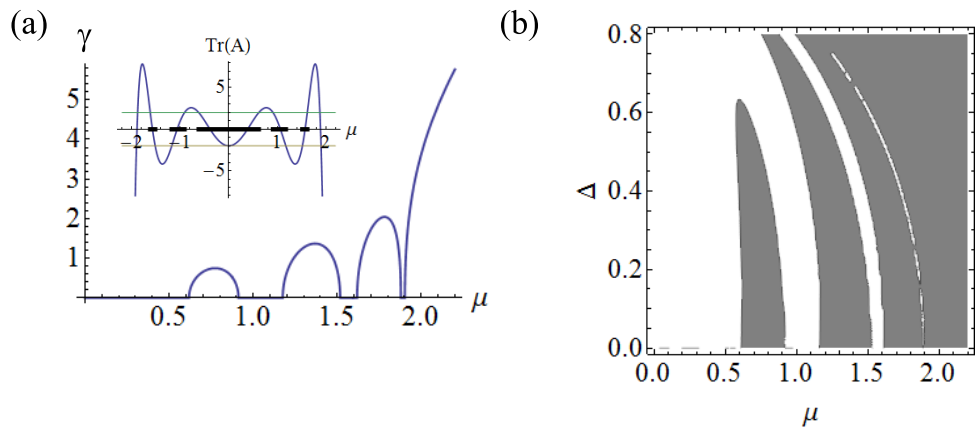,width=13cm}
\caption{(a) Plot of the Lyapunov exponent $\gamma(\mu)$ for a period 10
potential with $\mu_n = \mu $ for $n = 0$ mod 10 and $-\mu$ otherwise.
Inset shows Tr $\mathcal{A}$. The allowed plane
wave states (corresponding to $\eta(\mu) = 0$) have $| \mbox{Tr }
\mathcal{A}| \le 2$. (b) The $\mu$-$\De$ topological phase diagram
(topological region indicated in white, non-topological region in gray).
Note that the values of $\mu$ for which $\gamma(\mu) = 0$ in (a) are precisely
those regions which become topological for arbitrarily small $\De$. This
relationship is elucidated in Sec. VI B.} \label{fig:periodic1}
\end{figure} \end{center}

\subsection{Ultra-long Period and Quasiperiodic Potentials}

\begin{center}
\begin{figure}
\epsfig{figure=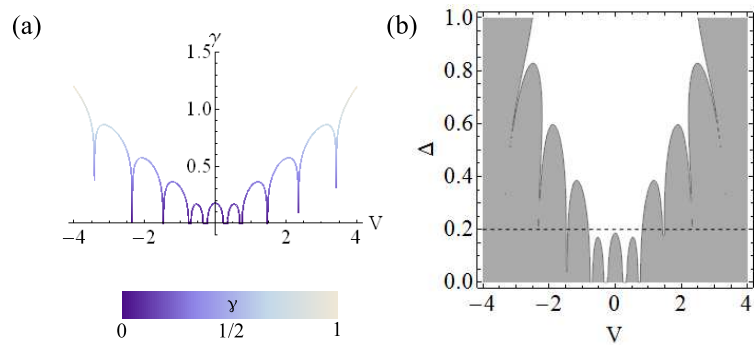,width=10cm}
\caption{(a) The Lyapunov exponent $\gamma$ of the normal state ($\De =0$)
for $\mu_n = V + 2 \cos \left( 2 \pi \om n \right)$ with $\om = 1/10$.
(b) Topological phase boundary showing the merging of the topological regions
as described in the text. For $\De \ll 1$, there are are 10 distinct
regions which are topological. At $\De = 0.2$, the four central
regions have merged to form a single region.} \label{fig:lyap}
\end{figure}
\end{center}

Building on our work in Sec. VI A, we consider periodic potentials of the
form $\mu_n = B \cos \left( 2 \pi \om n - \theta \right)$ but now allow
the period of the potential ($q$, for $\om = 2 \pi p/q$, for $p,q \in
\mathbb{N})$) to be either very long or infinite~\cite{tezuka,lang}. The
Hamiltonian formed from this potential is known as the \emph{almost Mathieu
operator}. For $B = 2$, the resultant equations of motion are known as
Harper's equations and arise in the context of a charged particle moving on
a 2D square lattice in the presence of an external magnetic
field~\cite{hofstadter}.

\begin{center}
\begin{figure}
\epsfig{figure=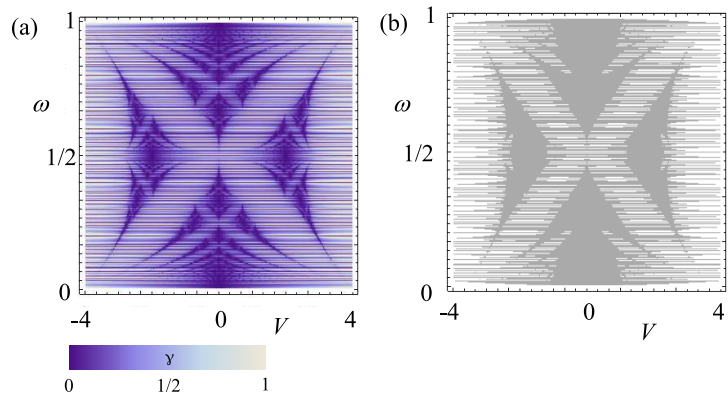,width=10cm}
\caption{(a) A colorscale plot of $\gamma (V,0)$ for
$\om = n/200$, $0 < n \leq 200$. Darker regions correspond to
smaller values of $\gamma$. The characteristic striations show the
spectrum's sensitivity to the period of the potential; i.e., small changes
in $\om$ can lead to large changes in the period (which is given by $q$,
where $\om = p/q$, with $p$ and $q$ relatively prime). The resulting
figure is reminiscent of the fractal known as Hofstadter's butterfly. (b) The
topological phase diagram for $\De = 1/5$ mimics the low-lying values of
$\gamma$ in (a).} \label{fig:butterfly}
\end{figure}
\end{center}

\begin{center}
\begin{figure}
\epsfig{figure=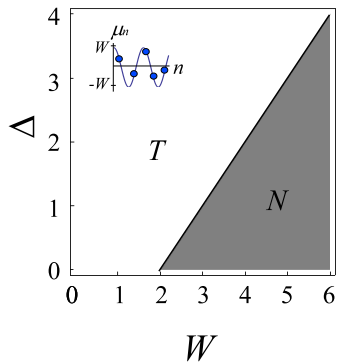,width=5cm}
\caption{Topological phase diagram for a potential $\mu_n = W \cos \left(
2 \pi \om n \right)$. The symbols $T$ and $N$ refer to topological and
non-topological regions respectively.} \label{fig:quasi}
\end{figure} \end{center}

For $\om$ irrational, the associated normal state energy spectrum is a fractal known as Hofstadter's
butterfly~\cite{hofstadter}. This fractal structure arises from a property of
the quantity Tr $\mathcal{A}$ (with $\De = 0$, see Sec. VI A) for the operator $\mu_n = 2 \cos \left(
2 \pi \om n - \theta \right)$ (see Eq.~\eqref{eq:matrixA}). This operator has the special property that for $\om = p/q, p,q \in \mathbb{N}$, there are $q$ distinct bands~\cite{hofstadter}. This is a special property of Harper's equation. For instance, note that in
Fig.~\ref{fig:periodic1}, although $q = 10$, only 5 distinct bands exist. In
contrast, Fig.~\ref{fig:lyap} shows that all $q = 10$ bands are present
for Harper's equation. For example, if we imagine letting $\om \rightarrow \sqrt{2}$
by taking a sequence of better and better rational approximations, the number of
bands will grow without bound. This is the origin of the unusual point-set topology
of the spectrum for irrational $\om$. A plot of the allowed bands as a
function of the energy and $\om$ leads to a fractal known as
\emph{Hofstadter's butterfly}~\cite{hofstadter}. The general features of
Hofstadter's butterfly maybe seen in Fig.~\ref{fig:butterfly}(a). Indeed,
for $\om$ irrational, the total `length' associated with the spectrum is zero
(by length, we mean $\int f(V) dV$, where
$f$ is 1 if $V$ corresponds to a point in the spectrum, 0 if not). Formally,
the Lebesgue measure of the almost Mathieu operator is $|4 - 2 |B |
|$~\cite{last}. Very roughly, as $q \to \infty$ for $B = 2$, the length of
the spectrum (as a function of energy) is zero, although there are still an
infinite number of distinct bands! These seemingly paradoxical properties are the
features of a so-called Cantor set.

These remarkable properties lead us to consider the topology of a 1D system subject to a
potential of the form
\begin{equation}
\mu_n = V + 2 \cos \left( 2 \pi p n /q \right).
\end{equation}
The allowed zero energy plane wave states in the $\om-V$ plane correspond to Hofstadter's butterfly. Of interest are the properties of the topological phase diagram as $\De$ is `turned on'. This problem is easily solved given the Lyapunov exponent $\gamma$ of Harper's equation which we have plotted in Fig.~\ref{fig:butterfly}(a). This plot is different from the usual plot of Hofstadter's butterfly in that it shows the value of the Lyapunov exponent $\gamma$ rather than just the spectrum (the spectrum corresponds to points for which $\gamma = 0$). Indeed, a feature of this plot are characteristic horizontal striations, which show the sensitivity of the localization length to the period of the potential ($q$).

We begin by considering the topological properties of a potential with a given $\om$. Fig.~\ref{fig:lyap} plots the Lyapunov exponent of a particular potential ($\om = 1/10$) as a function of $V$. As shown in Fig.~\ref{fig:lyap}(b), for the same potential and $\De \ll 1$, there are $q$ topological
regions inherited from the normal state. These distinct regions fuse as $\De$ is increased. The precise value of $\De$ for which two topological regions merge is determined by the strength of the Lyapunov exponent between the gaps. Since this quantity tends to be larger for larger $|V|$, the gaps closer to $V = 0$ tend to merge before those with larger $|V|$. We now generalize to the full $\om-V$ space. As expected from our general analysis, the
normal state properties (Fig.~\ref{fig:butterfly}(a) directly inform the
topological phase diagram (Fig.~\ref{fig:butterfly}(b). We see that as superconductivity is increased, Hofstadter's butterfly is `filled in' by non-topological
regions. The value of $\De$ at which the system becomes topological is extremely sensitive to the period $q$ of the potential.

The case of Harper's equation allows us to study the crossover exhibited from periodic to quasiperiodic potentials. We now consider the case $\mu_n = W \cos \left( 2 \pi \om n \right)$ with $\om$ being irrational. The normal state features of this potential have been well-studied; the system is metallic (i.e., there are plane wave states at $E = 0$) for $W < 2$ and exhibits a
metal-insulator transition at the critical value $W = 2$~\cite{svetlana}. The normal state Lyapunov exponent takes the form $\gamma (W,0) = \ln \left( W / 2 \right)$ for $W > 2$ and $0$ for $0 \leq W \leq 2$ for $\om$
irrational~\cite{delyon,andre}. Eq.~(\ref{eq:phaseboundary}) then predicts a
topological phase for
\beq \De > \frac{1}{2} W - 1. \label{eq:quasiperiodic} \eeq
This result holds for all values of $\De > 0$ given that the transformation
$\om \to \om + 1/2$ yields Eq.~(\ref{eq:duality}) and that the duality
transformation, $\De\to 1/\De$ and $W \to W/\De$, leaves
Eq.~(\ref{eq:quasiperiodic}) invariant. Finally, Eq.~(\ref{eq:specialpoint1})
also shows that the point $(W,\De) = (4,1)$ lies on the phase boundary.

\section{Disordered Potentials}

\begin{center} \begin{figure} \epsfig{figure=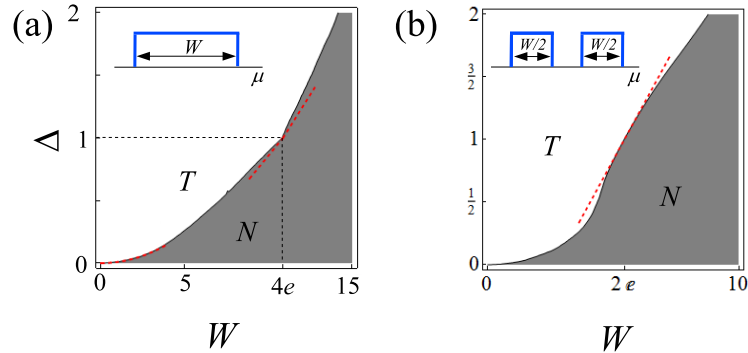,width=10cm}
\caption{Topological phase diagrams for (a) box and (b) double box potentials.
The red dotted lines indicate that at $\De =1$, the slope of the phase
boundary, $d\De/dW$, has a discontinuity in (a) but is continuous in (b).}
\label{fig:disorder1} \end{figure} \end{center}

The topic of disordered superconducting wires that break the SU(2) symmetry associated with spin, namely those of the symmetry classes
D and BDI discussed in previous sections, has been actively
researched for over a decade~\cite{altland,brouwer1,gruzberg,fidkowski1}. One of the highlighting features
of these systems is that their symmetry properties greatly alter localization physics. In particular, while Anderson localization dictates that states are always localized
in one-dimension in the presence of even the weakest disorder, these systems allow for the presence of
a critical disorder point in parameter space that permits a delocalized state at zero energy~\cite{altland,brouwer1,gruzberg,fidkowski1}. The point acts as a mobility edge in that it separates two localized regimes. In these systems, a variety of approaches have probed the manner in which the density of states diverges at zero energy, odd-even effects for coupled chains, and conduction
properties (since charge is not a conserved quantity for excitations about the superconducting
ground state, one studies thermal insulating/conducting properties).
In pioneering work by Motrunich et al.~\cite{motrunich}, it was shown that the localized phases
separated by the delocalized point are fundamentally different from one another. One phase supports end Majorana modes while the other does not, consistent with the classification of the phases as topological and non-topological, respectively, as in the disorder-free case.

Recent studies have focused on this feature of the presence of Majorana edge modes in the context
of disordered systems~\cite{shivamoggi,sau,brouwer1,brouwer2,brouwer3,beenakker,adagideli,degottardi}. Along these lines, here we exploit the transfer matrix
techniques that we have developed in previous subsections to pinpoint the conditions for the existence of
these end modes in a full range of models of disorder and disorder strength. The map made to normal systems in Sec. VI B allows us to borrow extensively from literature on Anderson localization
and perform a comprehensive study. We derive general features of
the phase boundary for disordered superconducting systems, including a characteristic discontinuity
of the phase boundary at the point associated with the random field Ising model ($\De = 1$). We also obtain topological phase diagrams for a variety of disorder distributions based on our mapping
to an extensive range of known results from Anderson localization studies of normal systems.

\subsection{Setup and General Results}

We consider Eq.~(\ref{ham0}) with $\mu$ replaced by a spatially dependent $\mu_n$. The $\mu_n$ is
drawn from some distribution and is uncorrelated, i.e. $\langle \mu_n \mu_{n'} \rangle = U
\delta_{n,n'}$ and $\langle \mu_n \rangle = 0$. The quantity $U$ is the standard deviation
of the disorder and thus characterizes its strength.

For weak disorder, the Lyapunov exponent may be obtained from perturbation theory for the normal state system~\cite{derrida,disorderreview} and is given by
\begin{equation}
\gamma(W,\De = 0) = \left( \frac{\Gamma(3/4)}{\Gamma(1/4)} \right)^2 U.
\end{equation}
Applying Eq.~(\ref{eq:phaseboundary}), we obtain the condition for the topological
phase
\beq \De > \left( \frac{\Gamma(3/4)}{\Gamma(1/4)} \right)^2 U
\approx 0.114 ~U. \label{eq:low} \eeq
This result may be compared to that of a continuum model based on the Dirac
equation, which gives a non-topological phase for $\De >
\frac{1}{8} U = 0.125 U$ (see Ref.~\cite{brouwer2}). For disorder
distributions that are symmetric around 0,
the self-duality condition $P=\tilde{P}$ in Eq.~(\ref{eq:duality}) is
satisfied. In this case, we can employ this duality transformation,
$\De \to 1/\De$ and $U \to U/\De^2$, to show that Eq.~(\ref{eq:low})
also describes the phase boundary in the limit of \emph{strong} disorder. Finally, we mention
that near the Ising point ($\De = 1$), the disorder may give rise to a discontinuity in the phase diagram. We describe the physics of this phenomenon in the examples given now.

\subsection{`Box' Disorder}

As a generic representative for disorder, we now turn to the case of
`box' disorder for which the probability of $\mu_n$ falling at any point in
the range $-W/2 \leq \mu_n \leq W/2$ is equally likely. The low-energy
behavior as shown in the numerical simulation in Fig.~\ref{fig:disorder1}(a)
is in good agreement with Eq.~(\ref{eq:low}) (for box disorder, $U = W^2/12$).
Eq.~(\ref{eq:specialpoint1}) reveals that the phase boundary passes through
the point $(W,\De) = (W_c,1)$, where $W_c = 4e \approx 10.873$ (box
disorder) with $e$ being the base of the natural logarithm.

\textbf{Discontinuity of phase boundary}\--- A noteworthy find
is the observed discontinuity suffered by the phase boundary as it passes
through the random transverse Ising field point $\De/w = 1$ discussed in the previous
section (Figs.~\ref{fig:disorder1}(a) and
\ref{fig:disorder2}).
We can calculate the phase boundary near $\De/w \approx 1$ by noting that according to Sec. \ref{sec:mapping},
$\gamma \left( \frac{W}{\sqrt{1-\De^2}},0 \right)$, the effective Lyapunov
exponent that we seek in Eq.~(\ref{eq:phaseboundary}), corresponds to that of
very strong disorder for $\De \to 1$. In this limit, we can use the known
form of the normal state Lyapunov exponent for $W \gg 1$
\bea \gamma(W) &\sim& \ln \left(W/2 + \sqrt{W^2/4-4 }\right) -
\sqrt{W^2/4 - 4} - \ln 2, \non \\
&\sim& \ln \left(W / 2 \right) - 1 + 4/W^2 + \mathcal{O}\left(1/W^4\right),
\label{eq:strongdisorder} \eea
(there is a typo in the expression given in ~\cite{disorderreview}).
Substituting this expression into Eq.~(\ref{eq:phaseboundary}) and invoking
self-duality, we obtain the phase boundary (to linear order around $(W,\De)
= (4e, 1)$))
\beq \De \approx \begin{cases}
\frac{e}{2e^2 + 2}W-\frac{e^2-1}{e^2+1} \ \ \mbox{for } \De \le 1, \\
\frac{e}{2e^2-2}W-\frac{e^2+1}{e^2-1} \ \ \mbox{for } \De \ge 1.
\end{cases} \label{eq:ansatz} \eeq
As seen in Fig.~\ref{fig:disorder1}(a), this result is in reasonable agreement
with numerical simulations.

In order to understand the origin of this discontinuity, we investigate the phase boundary by performing perturbation theory in the quantity $(1-\De^2)$. A straightforward calculation shows that the eigenvalues of the matrix $\mathcal{A}$ are 0 and
\beq \la = \frac{1}{(1+\De)^N} \prod_{n = 1}^N \mu_n \left[ 1 + \left(
1-\De^2 \right) \sum_m \frac{1}{\mu_m \mu_{m+1}} + \left(1 - \De^2 \right)^2
\sum_{m < m'} \frac{1}{\mu_m \mu_{m+1} \mu_{m'} \mu_{m'+ 1}}+ ... \right].
\eeq
From this expression, it is straightforward to obtain the Lyapunov exponent in terms of the disorder distribution. Letting $\De = 1 + \delta$, we get
\beq \gamma = \langle \ln \mu \rangle - \ln 2 - \frac{1}{2}
\left(1 + 4 \langle 1/\mu \rangle^2 \right) \delta \eeq
up to order $\de^2$. Let us define $\tilde{\delta} = \frac{1-\De}{1+\De}$; we then get that the phase boundary for $\tilde{\delta} \approx 0$ obeys
\beq \langle \ln | \mu_n | \rangle = \ln 2 - \left( 1 + 4 \langle 1 / \mu
\rangle^2 \right) \tilde{\delta} + \mathcal{O}(\tilde{\delta}^2).
\label{eq:phaseboundarnear1}
\eeq
This shows that the phase boundary is fragile towards
singularities when $\mu_n$ is allowed to come arbitrarily close to zero.
Indeed, our simulations have shown that the discontinuity is absent for
disorder distributions that avoid zero energy. A large class
of disorder distributions that cover zero energy give rise to a discontinuity
in the slope of the topological phase boundary at $\De = 1$. This is another manifestation of the sensitivity that the phase diagram shows to $\mu_n$'s which are equal to zero (for example, see Eq.~\eqref{eq:specialpoint1} or Fig. 7).

\subsection{Other forms of disorder}

\textbf{`Double Box' Disorder}\---
 A useful test of our hypothesis on the fragility of the phase boundary for arbitrarily small values of $\mu_n$ is to examine a system
 for which the disorder distribution excludes $\mu_n = 0$ and thus we expect that the phase boundary be continuous near $\De = 1$.
 Consider the following distribution for the local chemical potential, $\mu_n$:
\beq f_{DB}(x;W) =
\begin{cases}
\frac{2}{W} \ \ \mbox{for } W/4 \leq | x | \leq W/2, \\
0 \ \ \mbox{otherwise}.
\end{cases} \eeq
Here, $W_c = 2e \approx 5.437$. Note that all the $\langle 1/\mu^n \rangle$ exist. We can find the behavior of the phase diagram near $\De = 1$ using Eq.~\eqref{eq:phaseboundarnear1}. More directly however, we note that to linear order near $\De = 1$, the only expression which is invariant under
Eq.~\eqref{eq:duality} is
\beq \De = \alpha W - 1. \eeq
This expression well describes the phase boundary for $\alpha = 1/e$. This phase diagram is shown in Fig.~\ref{fig:disorder1}(b) and is in good agreement with this prediction, in particular, being devoid of the singularity at $\De=1$.

\begin{center} \begin{figure} \epsfig{figure=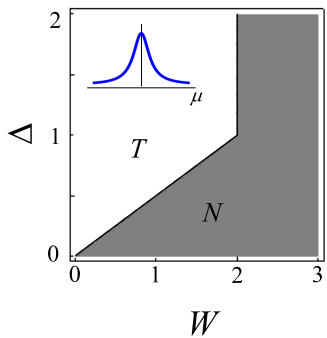,width=5cm}
\caption{Topological phase diagram for Lorentzian disorder.}
\label{fig:disorder2} \end{figure} \end{center}

\textbf{Lorentzian Disorder}\---
Finally, we turn to the specific case of disorder which is unbounded and has a diverging standard deviation, the
Lorentzian case. The distribution of local chemical potential is drawn from
a distribution of the form
\beq f_L(x; W) = \frac{1}{\pi} \frac{W}{x^2 + W^2}. \eeq
The phase diagram is exactly soluble in this case since the normal state
density of states is known exactly~\cite{lloyd}. The zero-energy Lyapunov
exponent, first obtained by Thouless~\cite{thoulessdos}, takes the
form $\gamma_L(W,0) = \ln \left( W/2 + \sqrt{1 + W^2/4} \right)$. Once again
invoking Eq.~(\ref{eq:phaseboundary}) and self-duality of the phase diagram
yields a phase boundary
\beq W = \begin{cases} 2 \De \ \ \mbox{for } \De \le 1, \\
2\ \ \mbox{for } \De \ge 1. \end{cases} \label{eq:lorentzian} \eeq
This result, as shown in Fig.~\ref{fig:disorder2}, is in excellent agreement
with numerical simulations. It should be pointed out that the features of this
phase diagram are extremely unusual. For instance, Eq.~(\ref{eq:low}) fails to
hold because the second moment $\langle \mu_n^2 \rangle$ is ill-defined
for $f_L$. This example is noteworthy since, for $W > 2$ the system is always
in a non-topological phase; no amount of $\De$ can drive the system topological.
Studying these examples has shown us, among other features, that typically the
larger the disorder, the more superconductivity is required for Majorana end
modes to exist, and that the topological phase diagram is highly sensitive to
the nature of the disorder distribution.

\begin{center}
\begin{figure}
\epsfig{figure=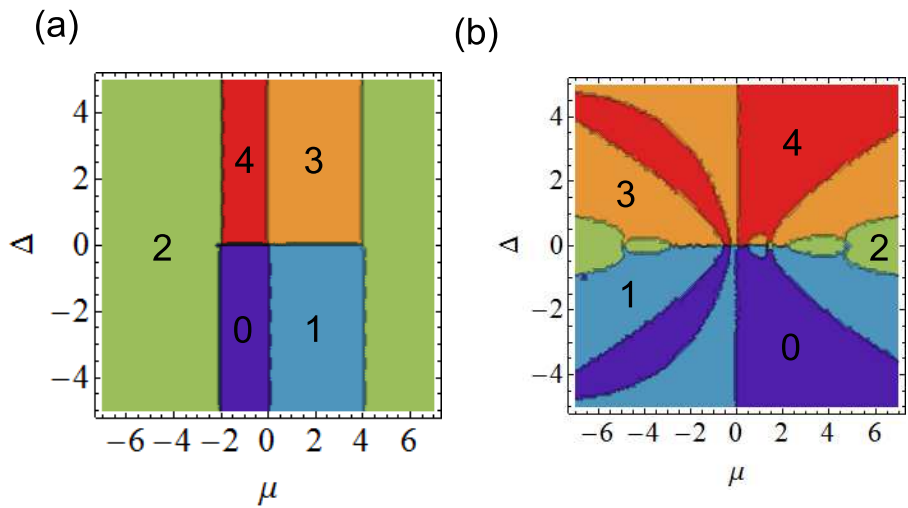,width=12cm}
\caption{(a) The topological phase diagram of a uniform class BDI system (for
which $\De$ is constrained to be real) described by Eq.~\eqref{ham4} with $J_0
= \mu/2$, $J_1 = J_2 = (w-\De)/2$, and $J_{-1} = J_{-2} = (w+\De)/2$, with
$w=1$. That is,
the system is similar to Eq.~\eqref{ham0} but with next-nearest neighbor
hopping and superconductivity equal to $w$ and $\De$. The different regions are
labeled according to the topologically protected quantity $n_f$, which reflects
the Majorana end mode structure as defined in Eq.~\eqref{eq:nf}. The phase
diagram was obtained using the method presented in
Sec.~\ref{sec:boundaryinvariants}. (b) The same phase diagram for a 15-site
system with a $\mu_n = \mu V_n$, where $V_n$ is a random real variable
between $-1$ and $1$. } \label{fig:multiple} \end{figure} \end{center}

We also mention that the method presented here may be applied to a Guassian distribution of disorder where a closed form of the Lyapunov exponent is known (see \cite{adagideli} for an analysis of this example). However, the result involves Airy functions and is quite complex and we omit this example given that it has the same qualitative behavior as the phase boundary for `box' disorder.

\textbf{Future Directions} \--- A very natural next step would be to consider spatially inhomogeneous potentials for a system in class BDI. The interplay between such potentials and phases exhibiting multiple Majorana modes should be quite rich. We offer a hint of this in Fig.~\ref{fig:multiple}. The figure indicates the intricate behavior which may occur in such cases, including re-entrant phases. Given this exciting behavior, we note that applied potentials offer a very promising route for the engineering of topological phase diagrams. We note however that the simple numerical methods which were applied here for a disordered system ($L\sim 10^4$) with nearest neighbor hopping will not work for systems with longer range hopping. This is due to the difficulty in extracting all the eigenvalues of a large product of matrices (and not just the largest one).

\section{Summary and Discussion}

In this work we have presented a comprehensive study towards enhancing
 our understanding of topological phases of one-dimensional superconducting systems,
TIs and Majorana end modes.
We began with reviewing a prototypical model, the Kitaev chain, which describes spinless
electrons hopping with an amplitude $w$ between the sites of a lattice,
$p$-wave superconducting order (denoted by a parameter $\De$), and an
on-site chemical potential $\mu$. This model has attracted a great deal of
attention as the simplest model which exhibits topological
phases and non-topological phases depending on the parameters appearing
in the Hamiltonian. In the topological phases, a long chain with open
boundary conditions has one zero energy mode localized at each end. The
non-topological phase has no end modes. The model can be mapped, using the
Jordan-Wigner transformation, to a spin-1/2 chain with $XY$ couplings and a
magnetic field applied along the $z$ direction. There are two topological
phases which have long-range order in the $x$ or $y$ components of the spin,
while the non-topological phase does not have any long-range order.

Next, we showed that there are a number of TIs which can
be used to characterize the different phases for uniform systems. There are different invariants
depending on whether the system is time-reversal symmetric or not, commonly
referred to as belonging to symmetry classes BDI and
D, respectively. The TIs can be defined by invoking the
bulk or the boundary properties for a large system with periodic and open boundary
conditions respectively. For time-reversal symmetric systems, the bulk invariant
is a $\mathbb{Z}$-valued winding number obtained from a closed curve lying in a
two-dimensional plane, formed by the set of Hamiltonians spanning momentum space.
 If time-reversal symmetry is broken,
the closed curve no longer lies in a plane and a winding number cannot be
defined. However, the Hamiltonians at the points $k=0$ and $\pi$ can be
used to define a $\mathbb{Z}_2$-valued invariant which can take values $\pm 1$.
As for boundary invariants, based on the Heisenberg equations of motion for open chains,
we obtained TIs that count the number of Majorana modes
(whose components we explicitly argued to be real)
at the ends of the system as well as their parity. We showed that in the
presence of time-reversal symmetry breaking terms, multiple Majorana modes,
while perhaps confined to the ends, disappear by obtaining complex components
and moving away from zero energy, thus allowing at most a single Majorana mode
at each end. We illustrated the bulk-boundary
correspondence which relates the values of the bulk and boundary invariants
to each other.

While the Hamiltonian for a Kitaev chain involves only nearest-neighbor hoppings
 and an on-site potential term, both only have interconnections between two types
of Majorana fermions, $a$ and $b$, we generalized
this model in several ways. First, we introduced long-range hoppings
of the Majorana fermions to construct time-reversal symmetric models which
can have any value of the $\mathbb{Z}$-valued TI in the bulk, and
therefore any number of Majorana end modes. For a time-reversal symmetric
system, we found that each end mode is purely of type $a$ or type $b$. As an
illustration of the kind of rich phase diagram that one may have with
long-range hoppings, we considered a particular model
in detail. This model has four terms in the Hamiltonian one of which
describes a next-nearest neighbor hopping. The model has four phases
with the number of end modes of types $a$ and $b$ varying from zero to 2.
The model can be mapped to a spin-1/2 chain; the corresponding model enjoys
a complete duality. All the four phases have long-range order in terms of
$x$ and $y$ components of either the original spins or the dual spins;
the phases are related pair-wise to each other through duality. Second, we
 introduced time-reversal breaking hoppings of Majorana fermions on
nearest-neighbor sites. We again obtained topological and non-topological phases
depending on the parameters. The topological phases have zero energy Majorana
end modes but now each end mode involves both $a$ and $b$ operators.
Interestingly, we found extended non-topological regions in parameter
space having a vanishing bulk gap. It is unusual to have such gapless
phases in parameter space; the other example of this that we are aware
of is the Kitaev model on the hexagonal lattice in two dimensions~\cite{kitaev2}.

Next, we explored a number of models in which the chemical potential
$\mu_n$ varies as a function of the site label $n$. We considered three
different situations in increasing order of complexity resulting in
a rich variety of phase diagrams. In order to find
these phase diagrams, we developed an important connection between
the transfer matrix of models with superconductivity and
the transfer matrix of non-superconducting models. Namely, we showed that
the localization length of the end modes of a model with $\De \ne 0$ can
be related to the Lyapunov exponent of a model with $\De = 0$. This connection
enabled us to make full use of the vast amount of literature available for
non-superconducting systems with complicated patterns of chemical potentials
to identify topological regimes when superconductivity is present.

The three kinds of patterns of $\mu_n$ that we analyzed are as follows.
First, we studied the case where $\mu_n$ is a periodic function of $n$
 commensurate with the lattice. We
found that the boundary between the topological and non-topological phases
depends on the period, amplitude and phase of $\mu_n$. We noted, interestingly,
that the boundary displays a cusp-like structure if $\mu_n$ happens to
vanish at certain sites in a periodic way. Second, we looked at
the case in which $\mu_n$ is quasiperiodic. Here we discovered a phase
diagram that has a fractal structure reminiscent of Hoftsadter's
butterfly (which appears in the energy spectrum of electrons moving on a
square lattice in the presence of a magnetic flux through each square which
is an irrational multiple of the flux quantum). Third, we explored the
case that $\mu_n$ is disordered and is drawn from some probability distributions.
Depending on the distribution and the value of the superconducting order
parameter $\De$, we find a wide variety of phase diagrams. If the disorder
distribution includes a range of values around $\mu_n = 0$, we found that
the phase boundary has a discontinuity in the slope at the point where $\De$
is equal to the hopping amplitude $w$. We showed this analytically using
perturbation theory in the parameter $\De /w - 1$ and we confirmed
this numerically for the box and Lorentzian distributions. In contrast, there
is no such discontinuity at $\De = w$ if the distribution does not allow
the possibility of $\mu_n = 0$, as we showed for the double box case.
(In order to facilitate these studies, we developed a duality relation
between regions with $\De/w > 1$ and $< 1$). The
relation between the superconducting system and the problem of Anderson localization
in normal electronic systems enabled us to analytically find the topological
phase diagram in some cases (the quasiperiodic pattern and Lorentzian
disorder), and to numerically find the phase diagram in other cases in
an efficient manner.

In the future we can consider many other generalizations of our work. As
mentioned at the end of Sec. VI, a very natural next step is to consider
multiple Majorana modes in disordered systems obeying time-reversal symmetry.
Additionally, we can study systems with spinful
electrons~\cite{altland,motrunich,brouwer1,gruzberg,fidkowski1}
and multi-channel wires~\cite{potter,fulga,stanescu1,tewari,gibertini,lim}.
Interactions between electrons are known to play a key role in the behavior
of one-dimensional systems, and it would be useful to study the effect of
interactions on the Majorana end
modes~\cite{ganga,sela,lutchyn2,lobos,fidkowski2} in the generalized
systems presented here.
Finally, the effects of finite temperature (through electron-phonon
interactions) and finite system length (which will lead to mixing between
the end modes) also need to be studied in detail.

In part, the interest in realizing Majorana fermions in a condensed matter setting
is motivated by the prospect of performing topological quantum computation by manipulating these particles~\cite{nayak}.
Recent theoretical work has shown~\cite{tjunction} that the manipulation of Majorana modes in a
T-junction geometry enacts unitary transformations on the (nearly) degenerate subspace formed from the Majoranas and the Dirac fermions they share. A crucial element of this proposal is the need to dynamically control
the topology of segments of the wire. In the case of a clean uniform wire (as was seen in Sec. II), the magnitude of the chemical potential alone controls the topology. Of particular relevance to such proposals, we have shown that
applied potentials allow the superconducting gap to serve as another `knob' for the topology, and thus an engineered potential could be used to exert more precise control over the topology of segments of the system. Many of the
theoretical tools introduced here may be applied to questions of the physics deep within the topological phase, not just at the phase boundary. Topological quantum computation requires a careful consideration of the coupling between different
Majorana bound states, crucial for the initialization and read-out steps in the protocol as well as understanding the effects of unwanted evolution between states in the ground state subspace~\cite{nayak}. The analysis carried out in Sec. VI to determine the topological phase boundary may be easily modified to find the localization length of the Majorana mode anywhere in the topological phase. This provides a means of estimating the coupling between Majoranas.

Finally, we turn to the possibility of testing our results in experimental
systems. The effectively
spinless models that we have studied in this paper are generalizations
of the experimental proposals of Refs.~\cite{lutchyn1,oreg} in the limit in
which the Zeeman energy greatly exceeds the superconducting gap and the
spin-orbit energy scale, and are applicable to the setting of the pioneering
experiment of Ref.~\cite{kouwenhoven}. This experimental setting thus provides a natural playground to explore
our generalizations, particularly in the the context of spatially varying
potentials. In principle, the predicted rich slew of topological phase diagrams
can be explored by a controlled application of periodic and disordered potentials.
The effects of time-reversal symmetry breaking can also be studied in a
controlled
fashion by the application of magnetic fields
A highly exciting prospect would be to discover phases with multiple independent Majorana
modes at the ends of a wire. While previous studies showed the existence of such phases in
multi-channel wires, perhaps potentially
realizable by fabricating a system involving several coupled wires, we explicitly
show here that they could exist within a single wire in the presence of long-range
hopping.

The detection and study of exotic particles was once thought to be the exclusive purview of
high energy physics. If the early experimental results are borne out, Majorana fermions give us yet another example of the synergy between quantum mechanics and many-body effects, rescrambling electrons whose identity is so commonly thought inviolable, into new and exotic objects. Our study of the numerous ways in
which Majorana modes can be realized in one-dimensional systems has been
inspired by the imminent possibility of discovering these fascinating objects.
In conclusion, we believe that the studies presented here can be further developed in interesting theoretical directions as well as investigated experimentally.

\section*{Acknowledgments}

For support, W.D. thanks the NSF under grant DMR 0644022-CAR and UChicago
Argonne, LLC, operator of Argonne National Laboratory, under contract No.
DE-AC02-06CH11357; M.T. thanks the CSIR, India; D.S. thanks DST, India
under Project No. SR/S2/JCB-44/2010; S.V. thanks the Simons Foundation under
Grant No.229047 and the U.S. Department of Energy, under Award No.
DE-FG02-07ER46453. W.D. and S.V. thank the Indian Institute of Science
for its generous hospitality at different stages of this work. We are
grateful to Ilya Gruzberg, Shinsei Ryu
and Jay Deep Sau for their illuminating comments.

\end{document}